\documentclass[]{aastex631}
\usepackage[utf8]{inputenc}
\usepackage[T1]{fontenc}

\usepackage{graphicx} 
\usepackage{xcolor}%
\usepackage[normalem]{ulem}
\usepackage{amsmath}

\definecolor{fleycolor}{RGB}{255,0,0}

\begin{document}

\title{The Role of Whistler and Ion Cyclotron Waves in Particle Escape from Mirror Modes in the Intracluster Medium} 

\author{Petr Ugarov}
\affiliation{Department of Physics, University of Wisconsin-Madison \\
Madison, WI 53706, USA}

\author{Francisco Ley}
\thanks{Present address: Universidad de Chile. email: fley.astro@gmail.com}
\affiliation{Department of Astronomy, University of Wisconsin-Madison \\
Madison, WI 53706, USA}

\author{Ellen G. Zweibel}
\affiliation{Department of Physics, University of Wisconsin-Madison \\
Madison, WI 53706, USA}
\affiliation{Department of Astronomy, University of Wisconsin-Madison \\
Madison, WI 53706, USA}

\begin{abstract}

Electron and ion-cyclotron waves are well known to exist in solar system plasmas but their existence and importance in galaxy clusters is an open question. 
Guided by numerical simulations, (\cite{Ley2023Secondary}) argued that whistlers (electron-cyclotron) and ion-cyclotron (IC) waves are generated by trapped particles in mirror modes in the nonlinear stages of the mirror instability under ICM conditions. Building on this work, we construct a novel particle propagation simulation of the ICM plasma based on the static electromagnetic field configuration from the fully kinetic particle-in-cell (PIC) simulation of the nonlinear mirror instability by (\cite{Ley2023Secondary}). We study how the trapping rate of particles is related to the secondary waves driven by mirror modes. We observe that secondary whistlers and IC waves enhance trapped particle escape from mirror modes. We measure the particle-wave scattering rate by whistlers and IC waves, demonstrate that the scattering rates and wave amplitudes follow the proportionality relation  expected from quasilinear theory, and show the existence of a significant correlation between scattering rates and the excitation of secondary instabilities. 

\end{abstract}

\section{Introduction} \label{sec:intro}

The study of the intracluster medium (ICM) in galaxy clusters is significant for its role in understanding galaxy formation and evolution, and it also serves as a natural laboratory for one of the most unique astrophysical plasmas. According to observational evidence, in the absence of heating sources, the ICM should rapidly lose energy by the emission of X-rays, leading to cooling flows that would enhance star formation in cluster centers (\cite{FabianNulsen1977, VoigtFabian2004}). However, observational studies of galaxy clusters show significantly less X-ray emission at lower temperatures than predicted by cooling flow models (\cite{Peterson2003}), suggesting the existence of a heating mechanism which counteracts radiative cooling.

A number of heating mechanisms have been proposed over the years from AGN heating to cosmic rays (see e.g. \cite{Zweibel2018} for a review). Particularly interesting is the mechanism of turbulent heating (\cite{Zhuravleva2014,Zweibel2018}). A number of challenges make this hypothesis difficult to analyze, namely the ICM's weakly collisional nature - in the sense of the typical mean free paths of particles being much larger than their gyroradii (\cite{Schekochihin2005}; \cite{Schekochihin2006}). The weakly collisional ICM plasma presents the difficulty that a fluid description is insufficient and full understanding of ICM physics requires a kinetic scale description. 

In a weakly collisional plasma like the ICM, Coulomb collisions are not able to drive the system to a thermodynamic equilibrium 
faster than the forces which drive it away. When turbulence is also present, as is the case of the ICM, a free energy source becomes available, and the plasma responds by redistributing this energy via kinetic plasma processes like wave-particle interactions. Particularly interesting for this work is the appearance of a pressure anisotropy, in the sense that the pressure perpendicular to the direction of the local ambient magnetic field becomes different from the pressure parallel to it. Macroscopically, pressure anisotropy manifests as an effective anisotropic viscosity, that can in principle heat the plasma, and it has been shown that this heating mechanism can be thermally stable in the context of the ICM (\cite{Lyutikov2007,Kunz2011}). 

\cite{Ley2023} have studied this heating mechanism through magnetic pumping using shearing box particle-in-cell (PIC) simulations. The magnetic pumping process describes how any turbulent motion that locally varies the magnetic field strength  causes  pressure anisotropy, which energizes the plasma via anisotropic viscosity. Particularly important for this mechanism to be effective is the excitation of plasma kinetic microinstabilities, which break the adiabaticity of magnetic pumping, allowing a net heating and an entropy increase (\cite{Kunz2011,Ley2023}).

The efficiency of this heating mechanism via magnetic pumping depends on the nature of the kinetic microinstabilities. The most relevant are the so called mirror modes (\cite{Chandrasekhar1958,RudakovSagdeev1961,SouthwoodKivelson1993,Pokhotelov2004}) and oblique firehose (\cite{HellingerMatsumoto2000}), although \cite{Ley2023} found that the so called ion-cyclotron (IC) and whistler instabilities can also influence the heating efficiency, via their interplay with the mirror instability. It was then hypothesized that the trapping effect 
that mirror modes naturally produce on ions and electrons as the instability develops, can excite secondary ion-cyclotron and electron-cyclotron (whistler) waves (\cite{Ley2023Secondary}), in an analogous way as the so called whistler lion roars can be excited within mirror modes in the Earth's magnetosheath (e.g. \cite{Kitamura2020,Jiang2022}). This trapping and scattering process plays an interesting role in regulating the ion and electron pressure anisotropy in the presence of mirror modes. In particular, the ion pressure anisotropy can be leveled at the ion-cyclotron instability threshold instead of the mirror instability threshold, which effectively increases the overall heating efficiency of the mechanism. In the case of the whistler instability, the presence of whistler waves can also influence the heat transport in the ICM (e.g. \cite{Yerger2025}). Therefore, these secondary IC and whistler instabilities become important in regulating the ion and electron pressure anisotropy evolution, and therefore affecting the ICM transport properties. Consequently, studying the interaction of these secondary waves with ions and electrons is relevant to understanding the macroscopic processes that occur in ICM.

In this work, we study in greater depth the mechanism of scattering by these secondary IC and whistler instabilities and their interaction with the trapping process by the mirror modes. We construct a static particle propagation simulation of plasma evolution in a shearing magnetic field using the electromagnetic fields from the simulations presented in (\cite{Ley2023Secondary}) which allows us to consider the direct relationship between mirror mode trapping and the secondary waves driven by mirror modes: ion-cyclotron and electron-cyclotron (whistler) waves. This paper is organized as follows: in Section~\ref{sec:simsetup}, we describe the simulation setup. In Section~\ref{sec:results}, we describe our results and methods of identifying trapped particles. We consider the effect of secondary instabilities by comparing simulations including electric fields and excluding electric fields. We identify a significant interaction between mirror modes and secondary instabilities. We quantify the scattering rate given these secondary instabilities and show that it is insensitive 
to the mass ratio of the simulation. Last, in Section~\ref{sec:conclusion}, we discuss the implications of our results and future work.

\section{Simulation Setup} \label{sec:simsetup}
The simulations are 2.5D with $\hat z$ the ignorable direction. Throughout the paper we adopt the following notation: subscripts $i$ and $e$ refer to ions and electrons, with other Roman letter subscripts on particle properties standing for either. Subscripts ``$\perp$" and "$\parallel$" denote directions perpendicular and parallel to the evolving background magnetic field. We use $\mu\equiv p_{\perp}^2/B$ for magnetic moment, $\omega_c$  for gyrofrequency, $R_L\equiv\sqrt{k_BT/m\omega_c^2}$ for gyroradius, and $\Delta P\equiv P_{\perp} - P_{\parallel}$ for pressure anisotropy. Wave electromagnetic fields are indicated by $\delta$, and the Gaussian system of units is used.

This paper involves the results of two simulations: an earlier relativistic Particle in Cell (PIC) simulation by (\cite{Ley2023Secondary}) using the relativistic, fully kinetic TRISTAN-MP code (\cite{Spitkovsky2005}, using  a continuously driven, shearing box (\cite{Riquelme2012}) and the particle propagation simulation that is the main result of this work. We will refer to the former specifically as the TRISTAN simulation.

In the TRISTAN simulation, the magnetic field is initialized as a homogeneous field in the x-direction in a box with periodic boundary conditions. The box then undergoes a shearing motion $\textbf{v} = - sx\hat{y}$ for 
1.5 shear times $\tau_s \equiv s^{-1}$.
 
The shearing motion amplifies the magnetic field through  
induction such that 
\begin{equation}\label{eq:vecB}
\mathbf{B}(t)=B_0\left(\hat x +\hat y st \right)
\end{equation}
and $B(t) = B_0 \sqrt{1 + s^2t^2}$. Salient parameters of the simulations are given in Table \ref{table:TRISTANSimulations}.
\begin{table}[h]
\centering
\begin{tabular}{cccccc}
\hline \hline
\multicolumn{1}{l}{$\beta_i$} & \multicolumn{1}{l}{$m_i/m_e$} & \multicolumn{1}{l}{$\omega_{c,i}/s$} & \multicolumn{1}{l}{$k_BT/m_ic^2$} & \multicolumn{1}{l}{$N_{\text{ppc}}$} & \multicolumn{1}{l}{$L/R_{L,i}^{\text{init}}$} \\ \hline
20                            & 8                             & 800                                  & 0.02                              & 600                                  & 54                                            \\
20                            & 32                            & 800                                  & 0.01                              & 300                                  & 50                                            \\
20                            & 64                            & 800                                  & 0.01                              & 200                                  & 40                                            \\ \hline
\end{tabular}
\caption{TRISTAN simulations and their physical and numerical parameters. The physical parameters are the initial ion plasma beta $\beta_i=8\pi P_{i}^{\text{init}}/B_0^2$, where $P_{i}^{\text{init}}$ and $B_0$ are the initial ion pressure and magnetic field strength, the mass ratio $m_i/m_e$ between the ions and electrons, the magnetization parameter 
$\omega_{c,i}/s$, and the initial temperature normalized by the ion rest mass energy $k_BT/m_ic^2$. The numerical parameters are the number of particles per cell $N_{\text{ppc}}$, and the box size normalized by the initial ion Larmor radius $L/R_{L,i}$.}
\label{table:TRISTANSimulations}
\end{table}

\begin{figure*}
    \centering
    \begin{tabular}{c}
        \includegraphics[width=0.98\linewidth]{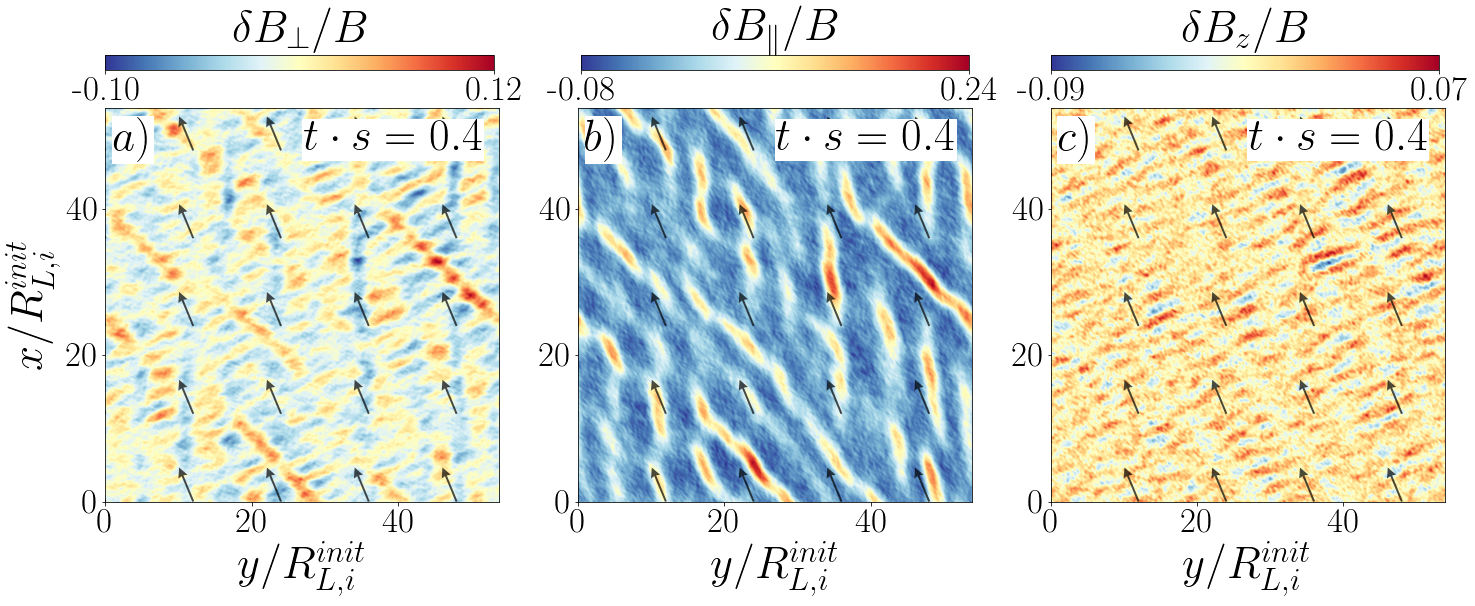}  \\
        \includegraphics[width=0.98\linewidth]{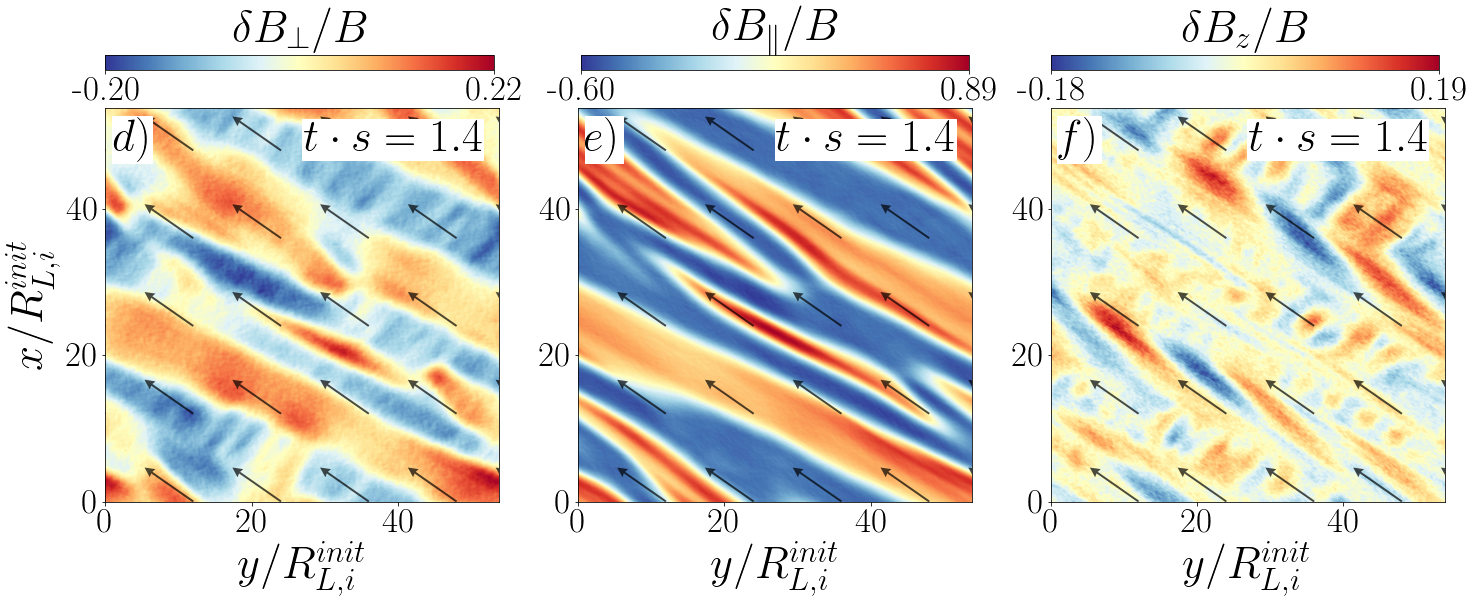}  \\
        \includegraphics[width=0.98\linewidth]{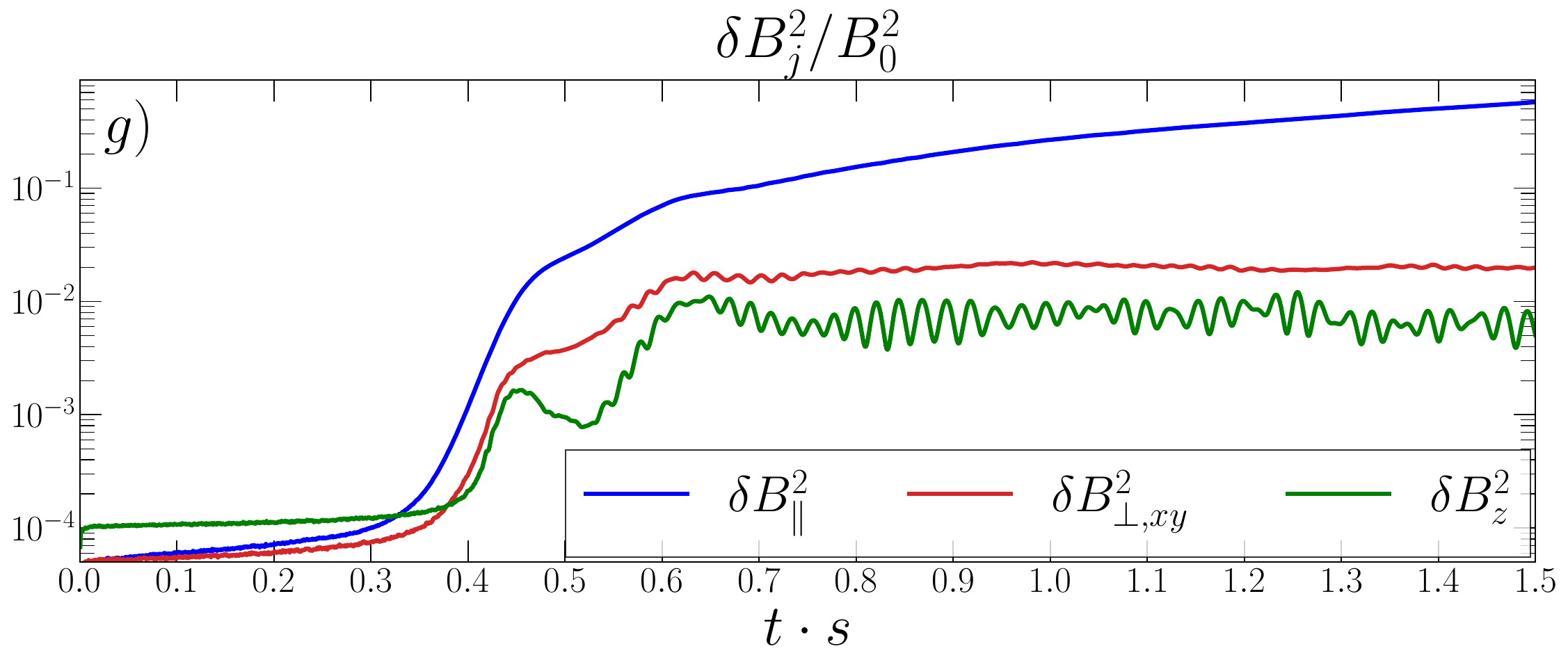} 
    \end{tabular}
    \caption{\textbf{First row:} The different component of magnetic fluctuations $\delta \textbf{B} = \textbf{B} - \langle\textbf{B}\rangle$ in the simulation domain for the TRISTAN simulation, at $t\cdot s=0.4$: $\delta B_{\perp}$ (Panel $a$) is the component perpendicular to the main field $\langle \textbf{B} \rangle$ in the $x$--$y$ plane of the simulation, $\delta B_{\parallel}$ (panel $b$) is the component parallel to $\langle \textbf{B} \rangle $ and $\delta B_z$ (panel $c$) is the component perpendicular to $\langle \textbf{B} \rangle$ in the direction out of the plane of the simulation. \textbf{Second row:} Panels $d$, $e$ and $f$ show the same as panels $a$, $b$ and $c$, but  but at $t\cdot s=1.4$. \textbf{Third row:} The evolution of the energy in the three component of the magnetic field fluctuations $\delta \textbf{B}$ normalized to $B(t)^2$, $\delta B_{\parallel}^2$ (blue line), $\delta B_{\perp,xy}^2$ (red line) and $\delta B_z^2$ (green line). This figure is adapted from \cite{Ley2023Secondary}.}
    \label{fig:fluctuations}
\end{figure*}

Without a scattering mechanism and provided that the magnetization parameter $\omega_{ci}/s\gg 1$,
the magnetic moments $\mu_j$ 
of both ions and electrons are 
adiabatic invariants. 
This must drive an increase in the particles' perpendicular component of the momentum, leading to an increase in the pressure anisotropy $\Delta P_j$. 
This is described as double-adiabatic or CGL scaling 
(\cite{ChewGoldbergerLow1956}). 
Eventually, however, $\Delta P$ 
passes the mirror threshold, $\Delta P_i \gtrsim 1/\beta_i$ ($\beta_i = 8\pi P_i/B^2$), 
excites the mirror instability (\cite{RudakovSagdeev1961}, \cite{Chandrasekhar1958}, \cite{SouthwoodKivelson1993}). The growth of the mirror instability leads to localized areas of strong magnetic fields (see fig. \ref{fig:fluctuations}$b$ and \ref{fig:fluctuations}$e$) and thus can be tracked in the evolution of the magnetic field fluctuations $\delta \textbf{B} = \textbf{B} - \langle \textbf{B} \rangle$ (where $\langle \textbf{B} \rangle$ is given by Eq. \ref{eq:vecB}). 

Figure~\ref{fig:fluctuations} shows the three components of the magnetic field fluctuations $\delta\textbf{B}$ (panels \ref{fig:fluctuations}$a$ $-$ \ref{fig:fluctuations}$f$) and the evolution of the energy in the different components of the magnetic field fluctuations (panel \ref{fig:fluctuations}$g$) that develop as a result of the mirror instability in the TRISTAN simulation (\cite{Ley2023Secondary}). The growth of the mirror instability in particular can be  
tracked in the $\delta B_\parallel$ component of the fluctuations (as can be seen in the solid blue line in figure~\ref{fig:fluctuations}$g$) (\cite{Pokhotelov2004}). Mirror modes begin to develop with an initial linear exponential phase starting at around $t \cdot s \approx 0.35$. After an initial linear exponential growth, the parallel component of the fluctuations $\delta B_\parallel^2$ levels off at $t \cdot s \approx 0.45$, marking the beginning of the secular stage (\cite{Kunz2014}). At this secular stage, the mirror modes begin to trap particles with low parallel momentum $p_\parallel \approx 0$ in mirror modes (\cite{SouthwoodKivelson1993}). The growth in the secular stage slows down further at $t \cdot s \approx 0.6$, when whistler and IC waves are excited (\cite{Ley2023Secondary}) which is seen in the oscillatory nature of the perpendicular components of the magnetic field fluctuations, $\delta B_\perp$ and $\delta B_z$ (red and green line in fig. \ref{fig:fluctuations}$g$, respectively). At this point, the mirror instability reaches the saturated stage ($t\cdot s \approx 0.7$, see $\delta B_{\parallel}(t)$ in fig. \ref{fig:fluctuations}$g$). In this late stage of the mirror instability, particles begin to escape from mirror modes, and the global ion pressure anisotropy saturates. The focus of the present work is to better characterize the mechanism of escape from mirror modes for both ions and electrons, during the saturated stage of the mirror instability. The main physical and numerical parameters of the TRISTAN simulations used in this work are listed in table \ref{table:TRISTANSimulations}. For more details of the TRISTAN simulation's construction, see (\cite{Ley2023Secondary}). 

In the particle propagation simulation, 
we use a static electromagnetic field configuration from the TRISTAN simulation at a particular timestep. To maintain continuity with the TRISTAN simulation, we use TRISTAN-MP dimensionless units for this simulation. 
We generally use a time step of $1$ through the simulation unless otherwise stated. For reference, the 
Larmor gyroperiods for ions $T^{i}$ and electrons $T^{e}$  
 are 6556 and 819 time steps, respectively given the numerical parameters in the first line of Table \ref{table:TRISTANSimulations}. Thus, the Larmor gyroperiod for both ions and electrons is well resolved.

The algorithm for the particle propagation simulation is based on a sequence of 4 steps with the last 3 steps in a loop: \bigskip

1) Spawn $n$ particles with random positions and initialized with velocities drawn from a Maxwell–Jüttner distribution.

2) Interpolate the fields to each of the particles' positions.

3) Propagate the particles using the Boris method (\cite{Boris1970,Birdsall2018}).

4) Update the positions and velocities of the particles. \bigskip

The particle velocities are generated 
by the inverse transform sampling method. 
We closely follow the methodology of (\cite{Zenitani2015}). Field interpolation is needed as the fields are computed along a grid in TRISTAN-MP, so a linear interpolation procedure is required to obtain the fields at the particle's position. We use the same interpolation procedure as in TRISTAN-MP. The Boris method is chosen for numerical orbit integration as it possesses good energy stability and conserves phase space volume. This approach is similar to those used, e.g., for propagating cosmic rays in fully developed, MHD turbulence (\cite{Lemoine2023,Kempski2023}).

To reduce PIC noise, we filtered the fields in Fourier space using a Gaussian filter with a cutoff at one electron Larmor radius, filtering out high $k R_{L, e} \gtrsim 1$. These modes lie below the electron gyroscale, where the spectral power is dominated by numerical noise. The filter therefore suppresses unphysical sub–Larmor–scale power while preserving the physical whistler and ion cyclotron modes which are the subject of our study.

The fields are static in time and taken at a particular time step in the TRISTAN simulation. For the results below, unless otherwise stated, we use a simulation time of 1.5 $t \cdot s$, or 1.5 shear times, corresponding to the end of the TRISTAN simulation. This timestep corresponds to a physically interesting time when the mirror modes and secondary instabilities are fully developed in the TRISTAN simulation (\cite{Ley2023Secondary}).

Considering the fact that mirror modes have a very weak electric field associated to them, and therefore the main source of electric fields in the TRISTAN simulation are the secondary whistler/IC instabilities developed during mirror's secular stage (\cite{Ley2023Secondary}), we can run simulations including both magnetic and electric fields, and also including only magnetic fields, in order to isolate the effect of mirror modes. The latter choice is naturally unphysical, but chosen as it provides a good numerical experiment to isolate the effect of trapping by mirror modes. Then, in simulations with electric fields, we can study how secondary instabilities can interact with mirror modes and compare with the simulations with only magnetic fields. 

Note that for parallel-propagating whistler waves with dispersion relation $\omega = \omega_{ci}(k d_e)^2$, the ratio of electric to magnetic fluctuations 
$\delta E/\delta B \sim (v_A/c)\,k d_i$. Using the simulation parameters in Table~\ref{table:TRISTANSimulations}, the Alfv\'en speed satisfies $v_A/c = \sqrt{(2k_B T_i/m_i c^2)/\beta_i} \approx 0.045$. This value is significantly larger than expected in the ICM, where typical parameters imply $v_A/c \sim 10^{-4}$--$10^{-3}$ (e.g \cite{Parrish2008}). Consequently, the electric-field fluctuations — and the associated particle scattering —
are likely to be overestimated in our PIC simulations relative to realistic cluster conditions. Taken together, our propagation simulations which retain $\delta E$, and our simulations which entirely omit $\delta E$ may roughly bracket reality as well as elucidating the physical processes that underlie particle scattering.

It is also worth mentioning that using static snapshots of electric and magnetic fields from the TRISTAN simulation to study the dynamical evolution of particles through these fields is a choice made for practicality as well as simplicity. However, as previous works also mentioned (\cite{Kempski2023,Lemoine2023}), it turns out to be a good approximation in some relevant instances. In our case, it becomes useful during the saturated stage of the mirror instability, where the growth of mirror modes is much slower than in the linear and secular stages, and comparable with the shearing time $s^{-1}$. Additionally, in a real astrophysical setting, the scale separation parameter is $\omega_{c,i}/s \sim 10^{4}-10^{5}$. Consequently, the interaction of particles and mirror modes is such that the particles would effectively see mirror modes that are quasi-static in their evolution, compared to any relevant timescale of particles.

\section{Results} \label{sec:results}

\subsection{Trapping Criterion} \label{subsec: trapping}

In order to study the mechanism of particle trapping in mirror modes, it is necessary to develop a criterion allowing us to split particles into two populations of trapped and passing particles 
(\cite{Kunz2014}). Here we use a simple measure of particle trapping: the sign changes counting method.

\begin{figure*}[h!]
  \gridline{
    \fig{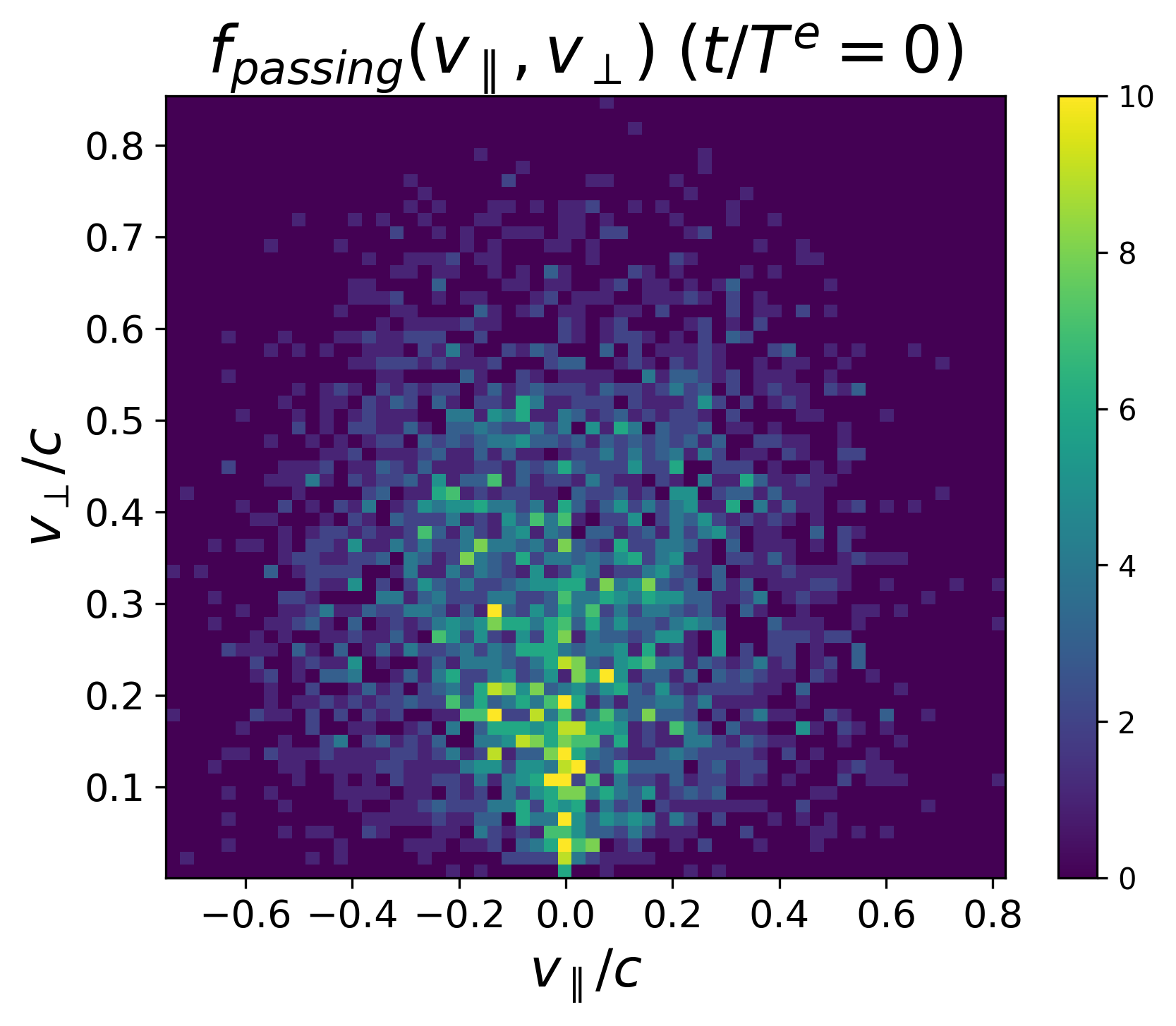}{0.34\textwidth}{}
    \fig{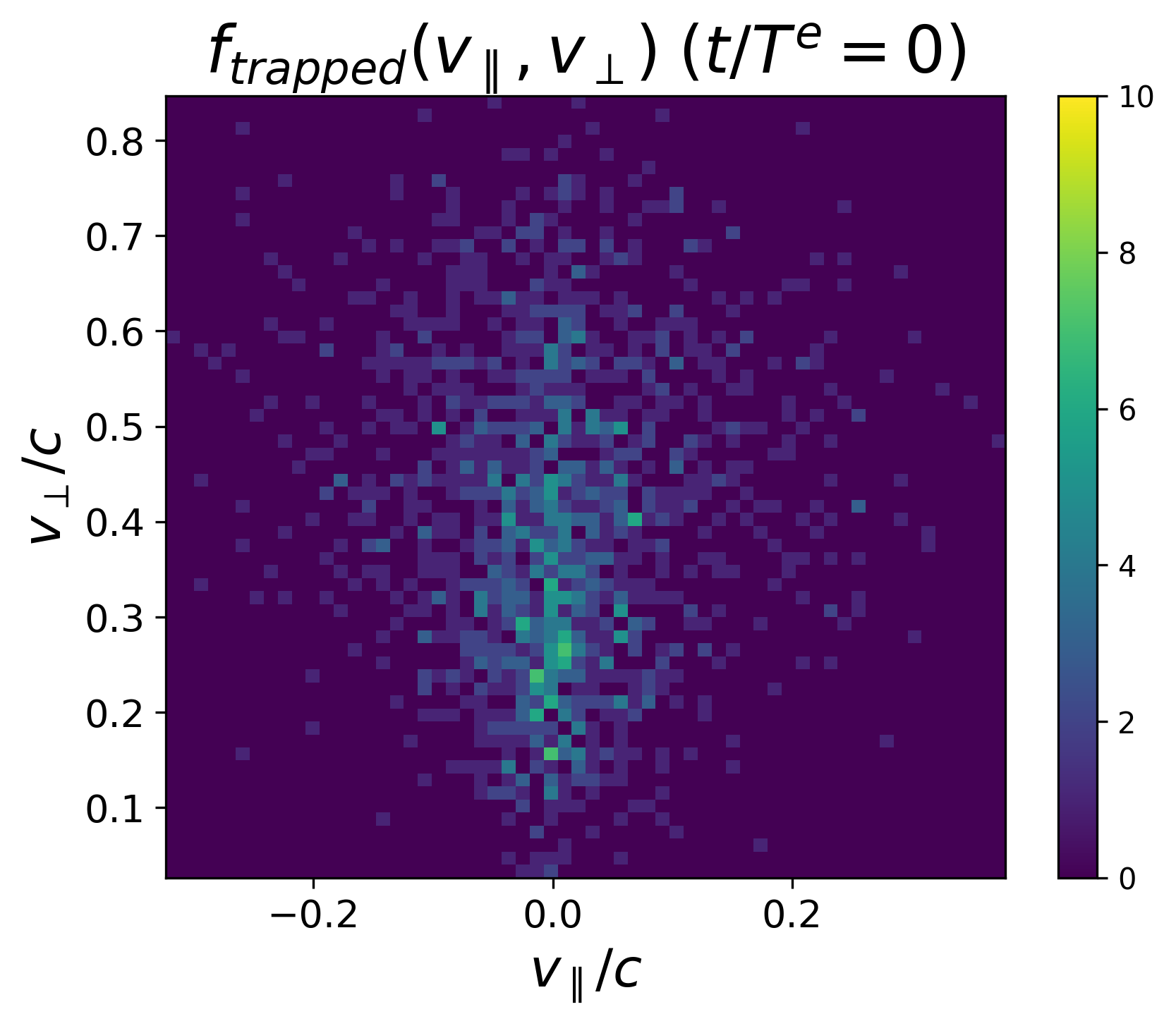}{0.34\textwidth}{}
  }
\vspace{-8ex}

  \gridline{
    \fig{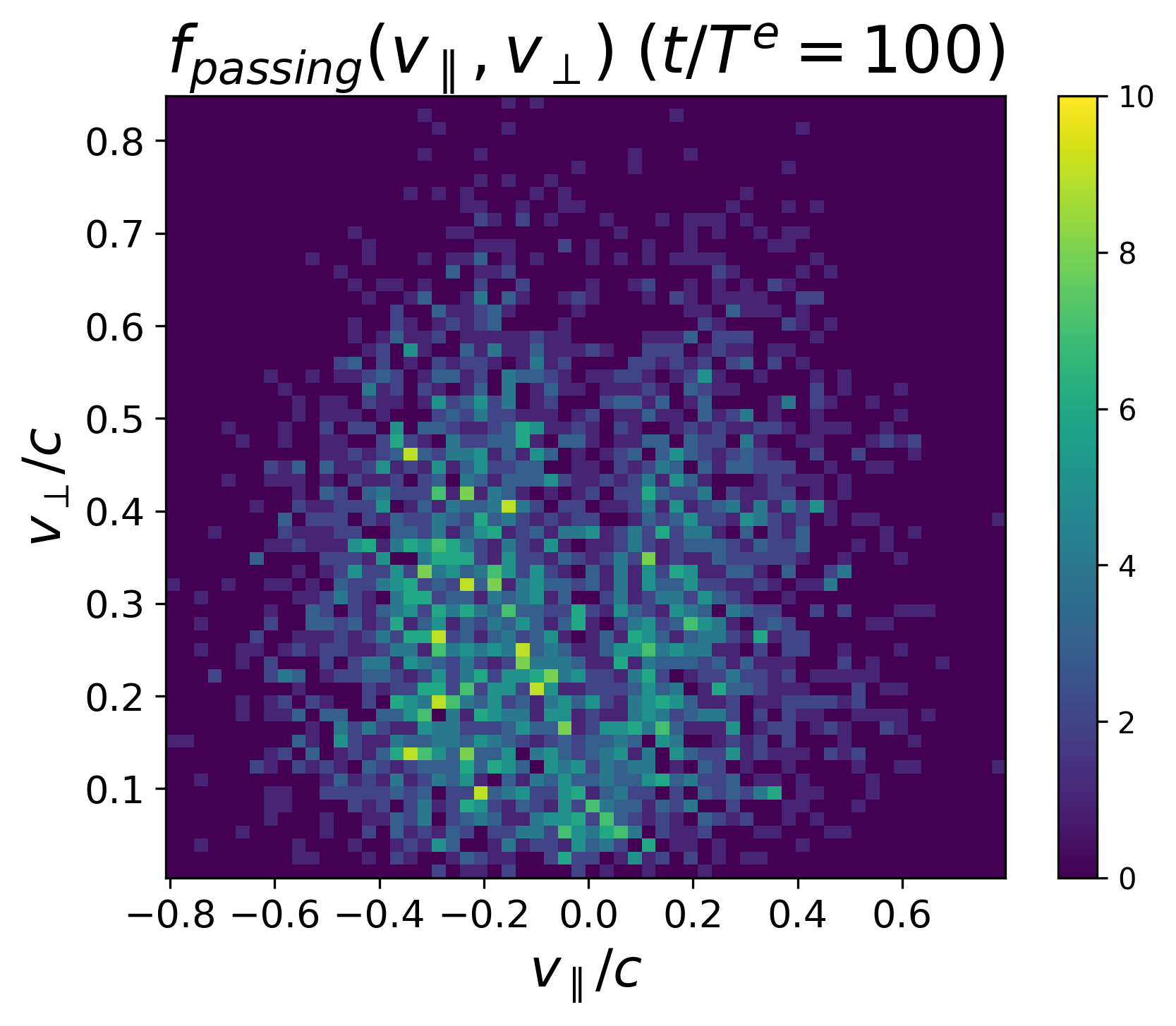}{0.34\textwidth}{}
    \fig{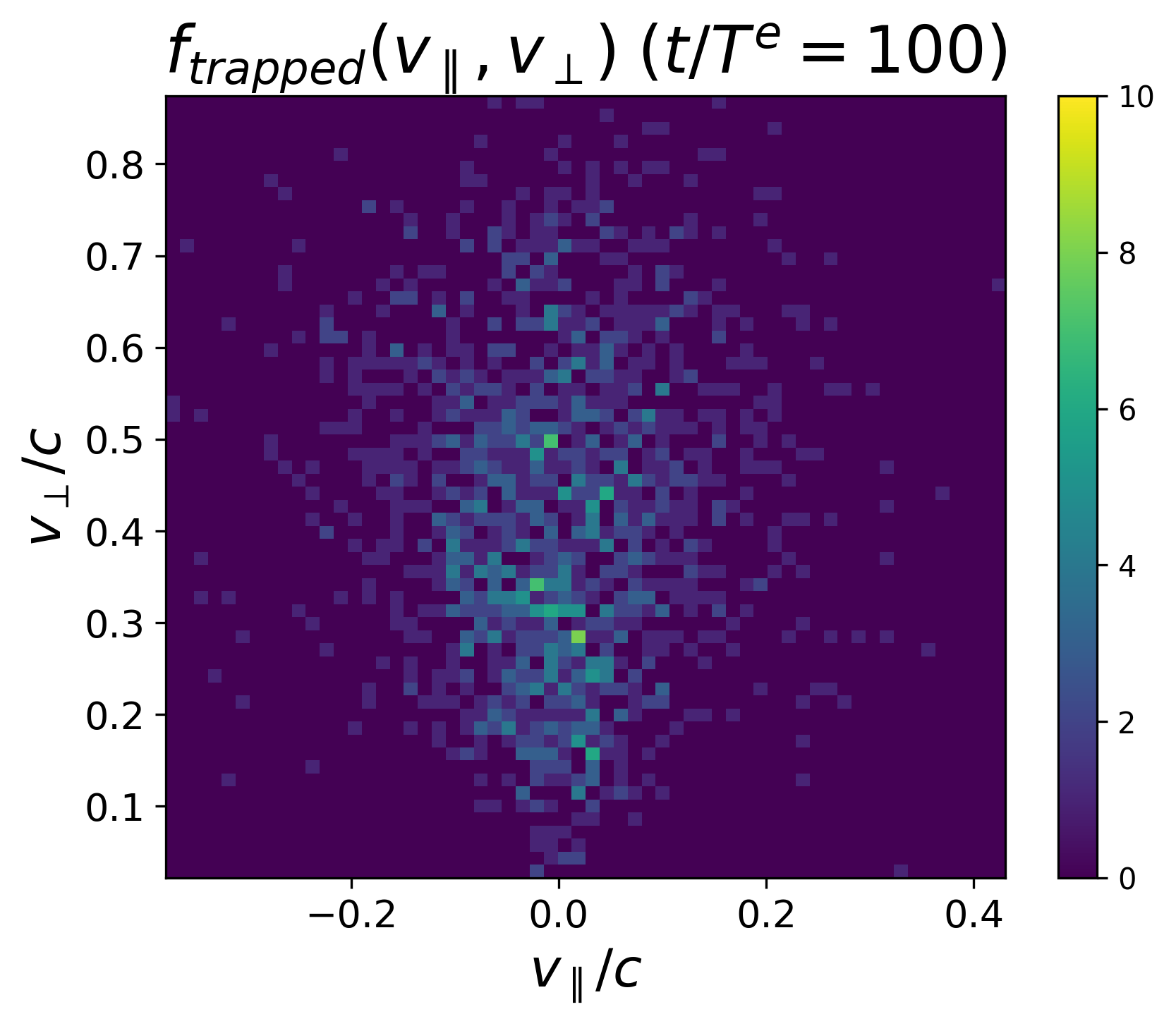}{0.34\textwidth}{}
  }

\vspace{-8ex}
  \gridline{
    \fig{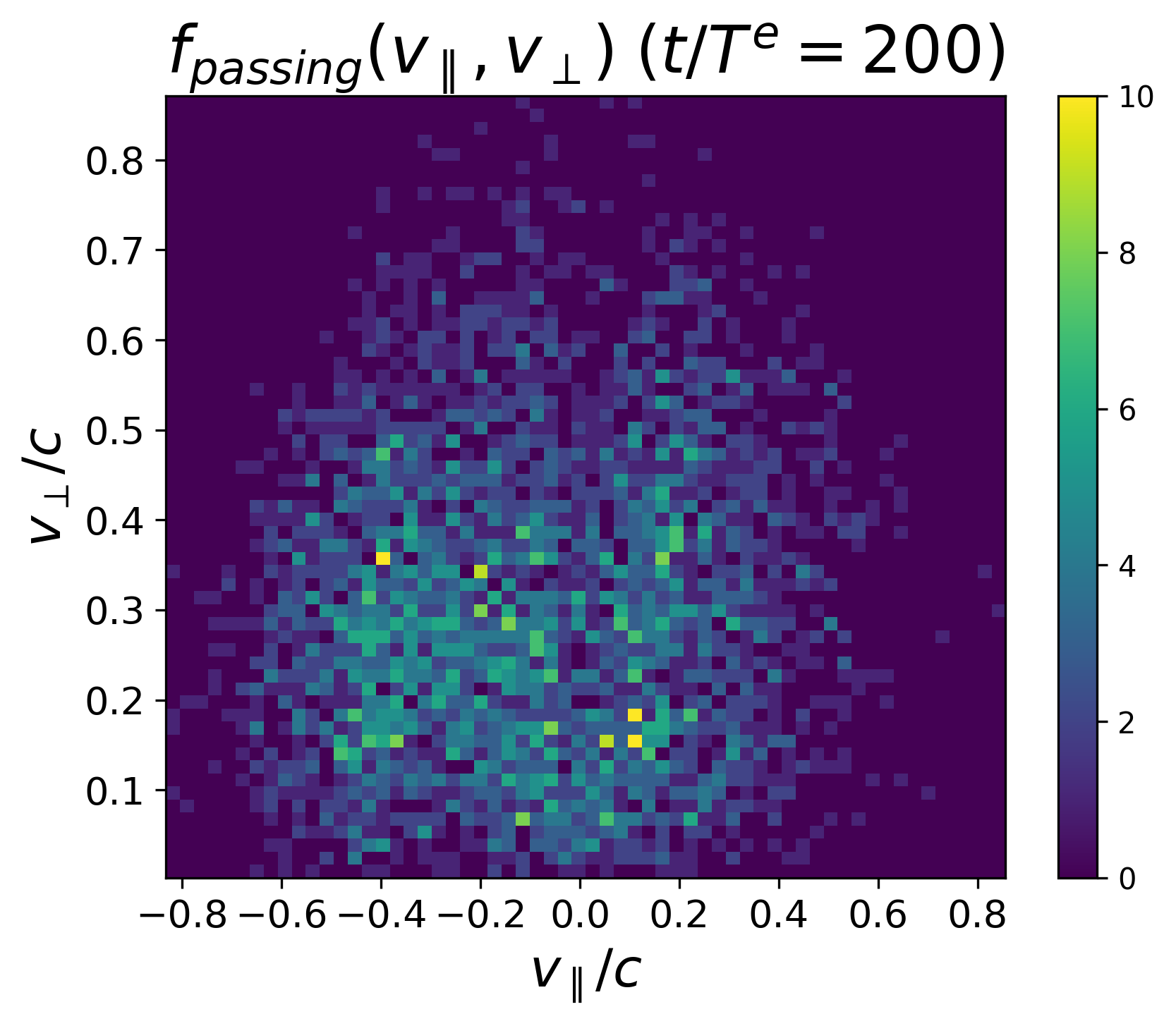}{0.34\textwidth}{}
    \fig{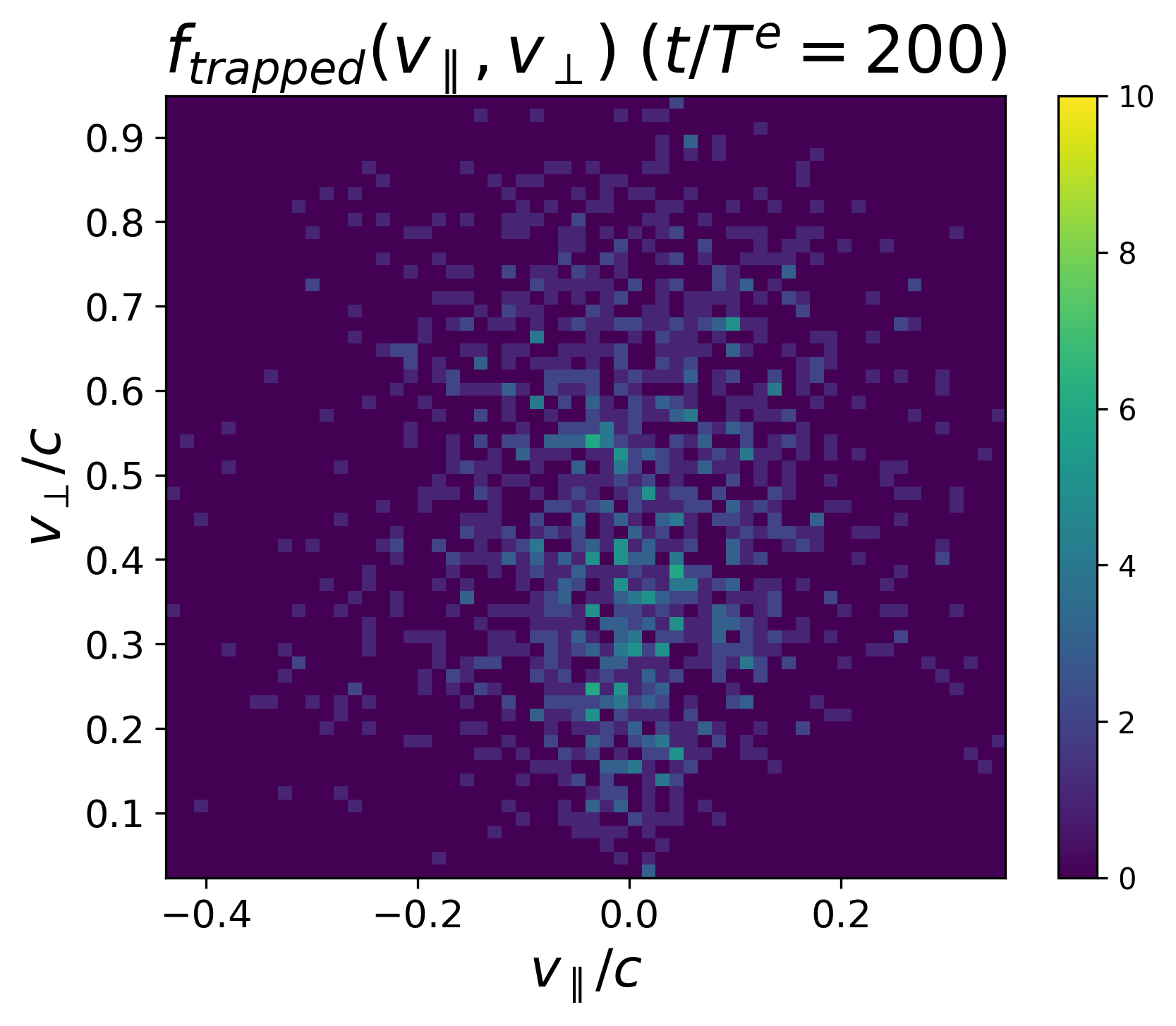}{0.34\textwidth}{}
  }

\vspace{-8ex}
  \gridline{
    \fig{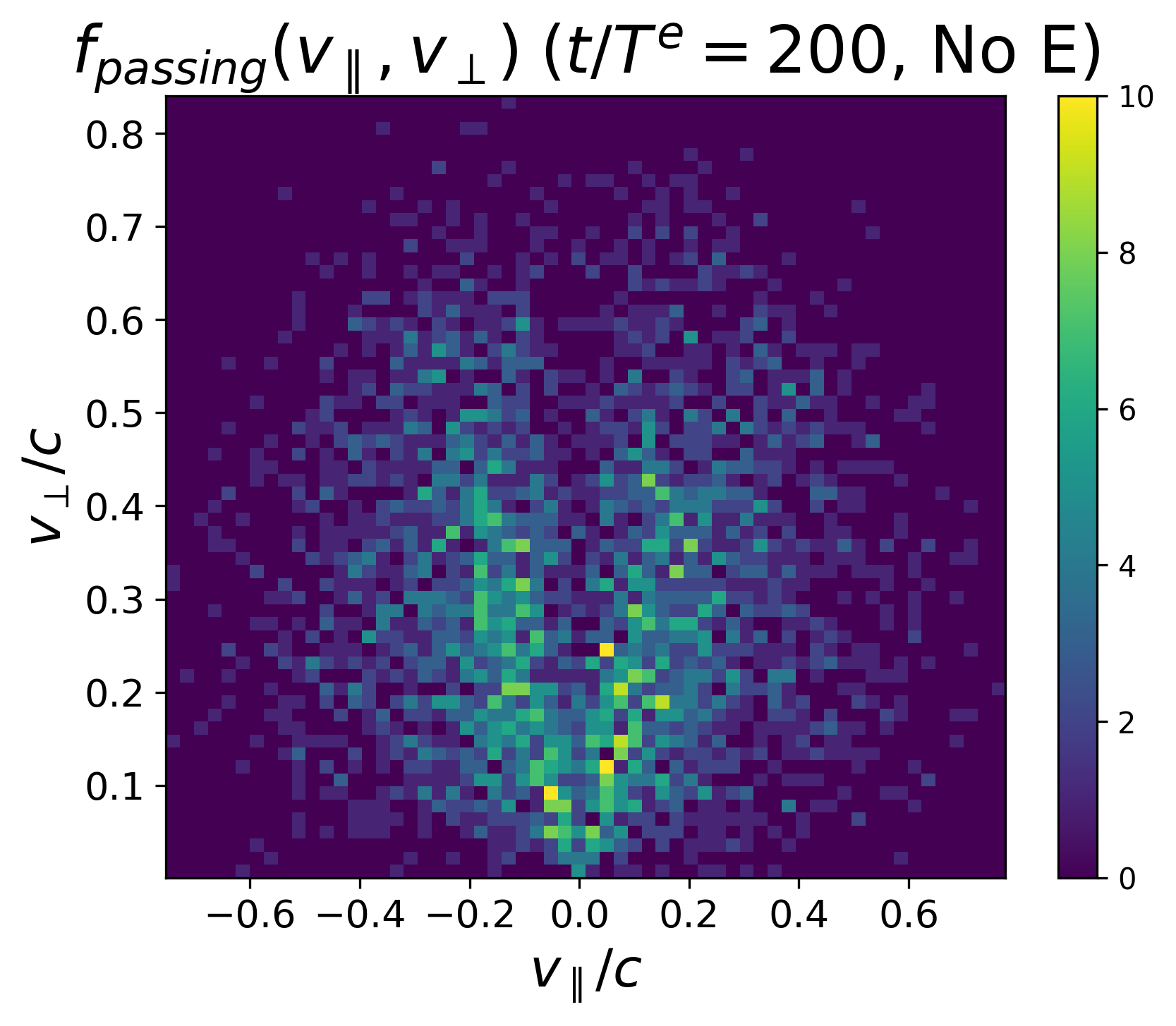}{0.34\textwidth}{}
    \fig{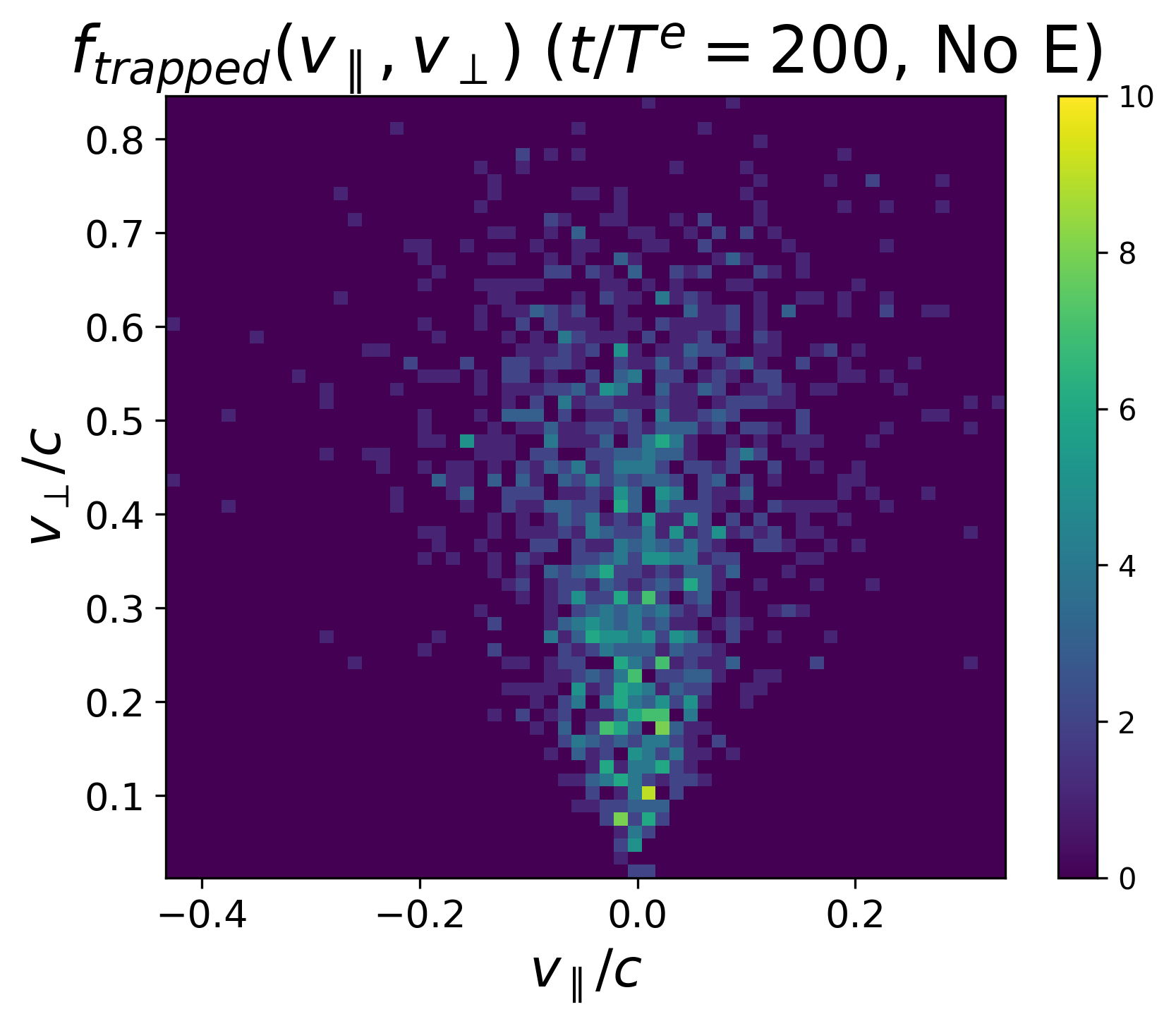}{0.34\textwidth}{}
  }
  \vspace{-5ex}
  \caption{
    Top three rows: The electron distribution functions $f(v_\parallel, v_\perp)$ for passing (left column) 
    and trapped particles (right column) in a particle tracking simulation including both magnetic and electric fields, 
    shown at the beginning $(t/T^e = 0)$, midway $(t/T^e = 100)$, and end of the integration $(t/T^e = 200)$. 
    Bottom row: Final electron distribution functions at $t/T^e = 200$ from simulations without electric fields. The concentration of trapped particles at lower $v_{\parallel}$ in comparison to that of the passing particles is obvious.
  }
  \label{fig:distfunctions}
\end{figure*}

The sign changes method is founded on the nature 
of particle trapping in magnetic mirrors: a magnetic mirror consists of a magnetic field volume which primarily varies along one axis, and is bounded by an increasing density of magnetic field lines at either end, forming a bottle-like shape. This creates a force parallel to the magnetic field, equal to $-\mu \nabla_\parallel \boldsymbol{B}$. 
In a static magnetic mirror, the magnetic moment and total energy are conserved (e.g.  \cite{Chen1984}). As the particle moves toward the end of the mirror, the magnetic field becomes stronger, thus to keep the magnetic moment constant, the perpendicular velocity $v_\perp$ must increase. However, since the total energy must be conserved, this then implies that $v_\parallel$ must decrease, until the particle is reflected towards the other side of the mirror. If a particle has too small of a perpendicular velocity or the magnetic field is too weak, the particle will escape the mirror trap. This can be quantified by the pitch angle, defined by $ \sin^2(\theta) = v^2_{\perp_0}/v_0^2,$ where the 0 subscript denotes initial values. The smallest pitch angle $\theta_m$ for which a particle can be trapped is equal to $\sin^2(\theta_m)= B_0/B_1 = 1/R_m,$ where $B_0$ is the magnetic field at the center of the mirror, $B_1$ is the field at the endpoints, and $R_m$ is defined as the mirror ratio (e.g.\cite{Chen1984}). This condition for a smallest pitch angle defines a cone-shaped boundary in $(v_\parallel, v_\perp)$ phase space. Particles with $v_\perp < \tan(\theta_m) |v_\parallel|$ escape, and form the region of phase space denoted as the loss cone. For the mirror modes in the TRISTAN simulations, we estimate $R_m \approx 1.097 \pm 0.047$ by fitting the edge of the loss cone in the trapped particle distributions in Figure~\ref{fig:distfunctions}. For this mirror ratio, we get a minimum $\theta_m \approx 72.7^{\circ}$.

With this picture in mind, particle trapping in an ideal magnetic mirror causes periodic behavior centered around zero in the parallel component of the particle's velocity $v_\parallel$, as the particle bounces back and forth between mirror points. As such, one simple measure of trapping is to count the number of sign changes that occur in a particle's $v_\parallel$ relative to the mean magnetic field. This acts as a statistical measure of trapping as the more sign changes a particle's $v_\parallel$ describes the more time it is likely to spend trapped. Sign changes can capture both short-term scattering events and longer-term trapping - for this reason, it is necessary to choose a sufficiently high threshold of sign changes to denote a particle as trapped. If a particle's sign changes are lower than this threshold, it is denoted as passing. 
In general, the threshold to select to separate the particles into 2 populations is arbitrary; in practice, we selected 50 as it is sufficiently high based on the following analysis of electron particle distribution functions and roughly divides the particles into 2 equal populations. This means that particles with less than or equal to $50 v_\parallel$ sign changes would be denoted as passing, and particles with more than $50 v_\parallel$ sign changes would be denoted as trapped.

\begin{figure}[h!]
\plottwo{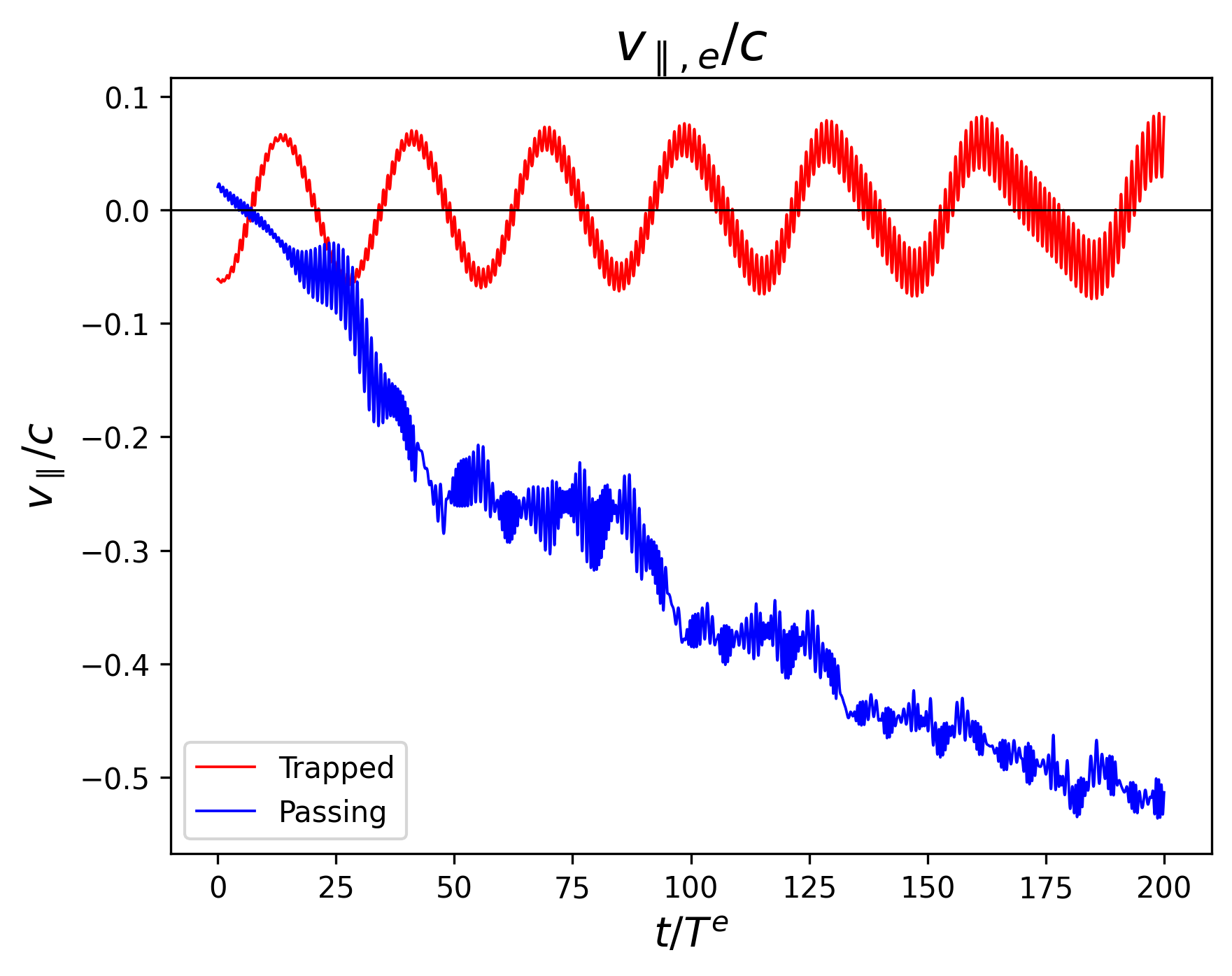}{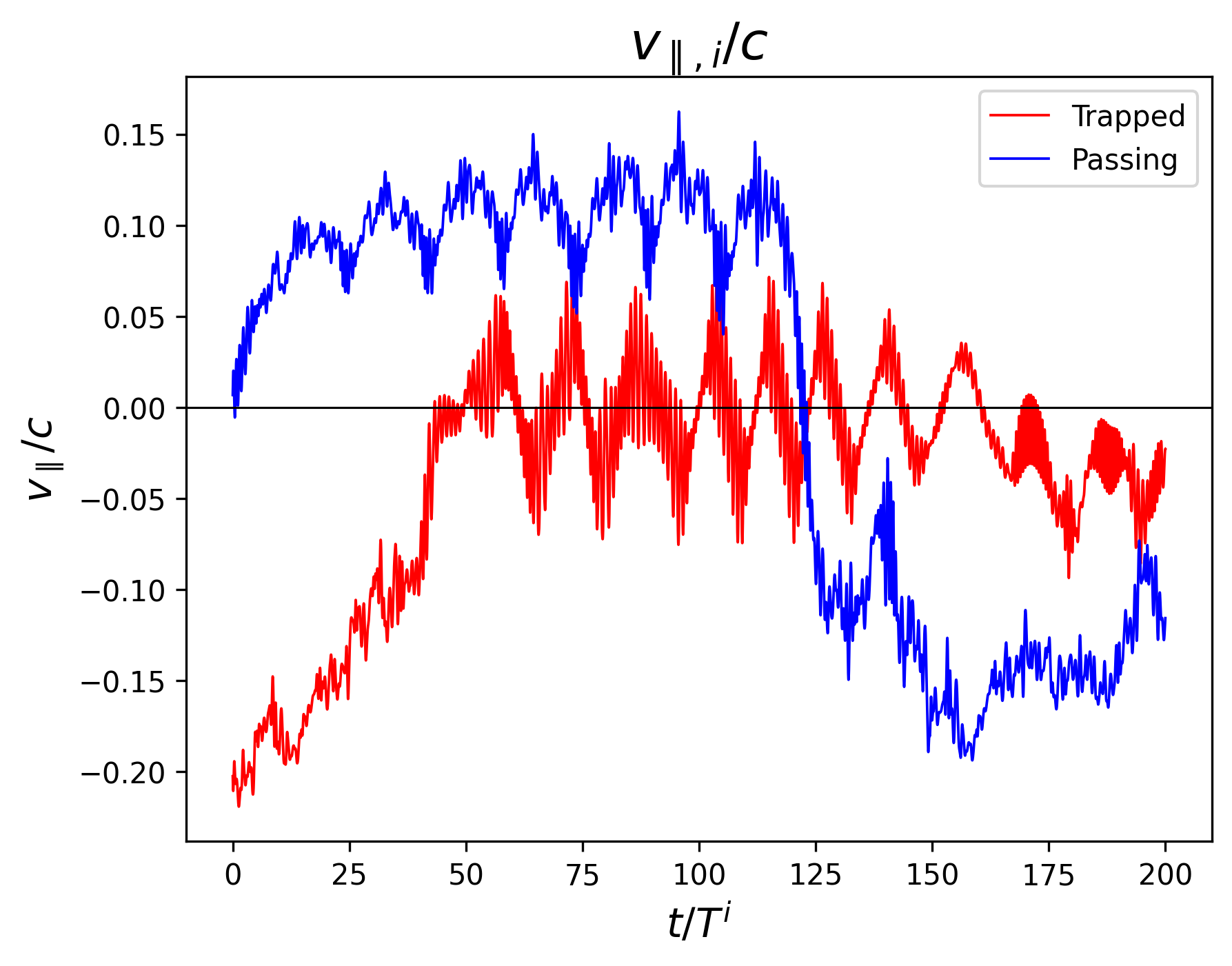}
\caption{Left: the parallel velocity $v_\parallel$ integrated over $200$ electron gyroperiods for a trapped and passing electron with 89 sign changes and 9 sign changes, respectively. Right: the parallel velocity $v_\parallel$ integrated over $200$ ion gyroperiods for a trapped and passing ion with 156 sign changes and 9 sign changes, respectively. 
\label{fig:RepresentativeIonsElectrons}}
\end{figure}

Figure~\ref{fig:RepresentativeIonsElectrons} shows the evolution of the parallel velocity $v_\parallel$ for representative passing and trapped electrons and ions as determined by the $50 v_\parallel$ sign changes threshold. In the left figure  corresponding to the evolution of the parallel velocity for electrons, we can see the characteristic oscillation of $v_\parallel$ for a representative electron with 89 sign changes. The parallel velocity oscillates around 0 for its entire evolution of 200 integrated electron gyroperiods, indicating that this representative electron was trapped in a mirror mode in this time span. In contrast, the passing electron with 9 signchanges does not have a similar oscillating pattern and never gets trapped in a mirror mode. In the right figure, we can see that the trapped ion with 156 sign changes exhibits the same oscillatory behavior most notably from around 50 to 150 integrated ion gyroperiods. The passing ion with only 9 sign changes does not exhibit the same oscillatory behavior around $v_\parallel = 0$. 

From this representative example we can see that trapped particles do not remain trapped through their entire evolution - it is only true that they are statistically more likely to be trapped in a given time interval than a passing particle. 

To further justify the choice of threshold, we considered the electron particle distribution $f(v_\parallel, v_\perp)$ at different timesteps in the integration for two populations classified by 50 sign changes: trapped and passing. The electron particle distributions for the trapped and passing populations are called $f_{\text{trapped}}(v_\parallel, v_\perp)$ and $f_{\text{passing}}(v_\parallel, v_\perp)$, respectively. Theoretically, in the limit of low scattering rate, $f_{\text{trapped}}(v_\parallel, v_\perp)$ would describe a loss cone distribution, as trapped particles in magnetic mirrors have $v_\perp \gg v_\parallel$. In Figure~\ref{fig:distfunctions}, we detail these results at different time steps of the particle's evolution.

In simulations without electric fields, the threshold of 50 sign changes produced a clear separation between the trapped and passing particle distributions. The trapped electron population (last row of Figure~\ref{fig:distfunctions}, right column) forms a loss cone distribution with $v_\perp \gg v_\parallel$. The passing population (left column) has $v_\parallel \gg v_\perp$ as expected. However, in the simulations where electric fields are present (the first three rows of Figure~\ref{fig:distfunctions}), the primary effect is increased dispersion in phase space due to the effect of scattering by secondary waves. In addition, a slight heating in the perpendicular direction can be seen. The effects of this spurious heating on our results will be discussed later in Appendix~\ref{sec:appendix}.

However, the particle dispersion in the simulations where electric fields are also present is more physically interesting, and it indicates the outsize role that electric field scattering plays in this simulation. In Section \ref{subsec:scatteringrate}, we will quantify this scattering rate from the wave-particle interactions.

\subsection{Scattering Rate} \label{subsec:scatteringrate}

As we described in the Introduction, particle trapping by mirror modes generates a positive pressure anisotropy, 
as trapped particles have very low $v_{\parallel,j}$ ($j=i,e$). This  anisotropy excites secondary instabilities which scatter the particles, reducing their anisotropy to  
a marginally stable state (\cite{Ley2023Secondary}). The relevant instabilities are ion-cyclotron (IC) waves  and whistler waves, which propagate through regions of low magnetic field strength inside mirror modes. As the secondary instabilities grow, they increasingly play a role in scattering the particles, the effect of which we can capture through an analysis of the scattering rate at different times in the simulation. Unlike the mirroring process, in which particles can preserve their magnetic moments, scattering by the  shorter wavelength IC and whistler waves (see Figure \ref{fig:fluctuations}) violates the ``slow change'' condition for adiabatic invariance and entails changes of $\mu$; this is useful for calculating the rate of scattering.

Here, we use a method for calculating the scattering rate utilized in (\cite{Zhou2023}) (See also e.g. \cite{Kunz2014, Kunz2020, Yerger2025}). We can assume that the scattering events are random and independent. Considering scattering as a Poisson process, we exponentially fit a histogram of the characteristic interval times $\tau_i$ that it takes electrons/ions to change their magnetic moment by a factor of $a = 0.5e$ over the entire integration of a particle. For each particle $i$, we record the intervals of time it takes for their magnetic moments to increase or decrease by a factor of $a$. This can be thought of as running a stopwatch for each particle while recording and resetting the time when the particle's magnetic moment significantly changes (a collision happens). The factor $a$ is arbitrary but has been chosen to be smaller than $e$ to increase statistical power (\cite{Kunz2020}). The slope $\langle \tau_{\text{coll}} \rangle$ of that linear fit of the histogram in log space gives an estimate of the average collision time due to the secondary instabilities in the simulation. Then, the effective collisionality $\nu_{\text{eff}}$ can be measured as $\nu_{\text{eff}} = \frac{\ln^2(0.5e)}{\langle \tau_{\text{coll}} \rangle}$, with the additional $\ln^2(0.5e)$ correction factor added to correct for the fact that $a=0.5e$, and not $e$ (\cite{Kunz2020}). The first bin of small characteristic times is not included in the linear fit to exclude particles encountering multiple field reversals during their gyromotion (\cite{Zhou2023}).

\begin{figure}[h!]
\plottwo{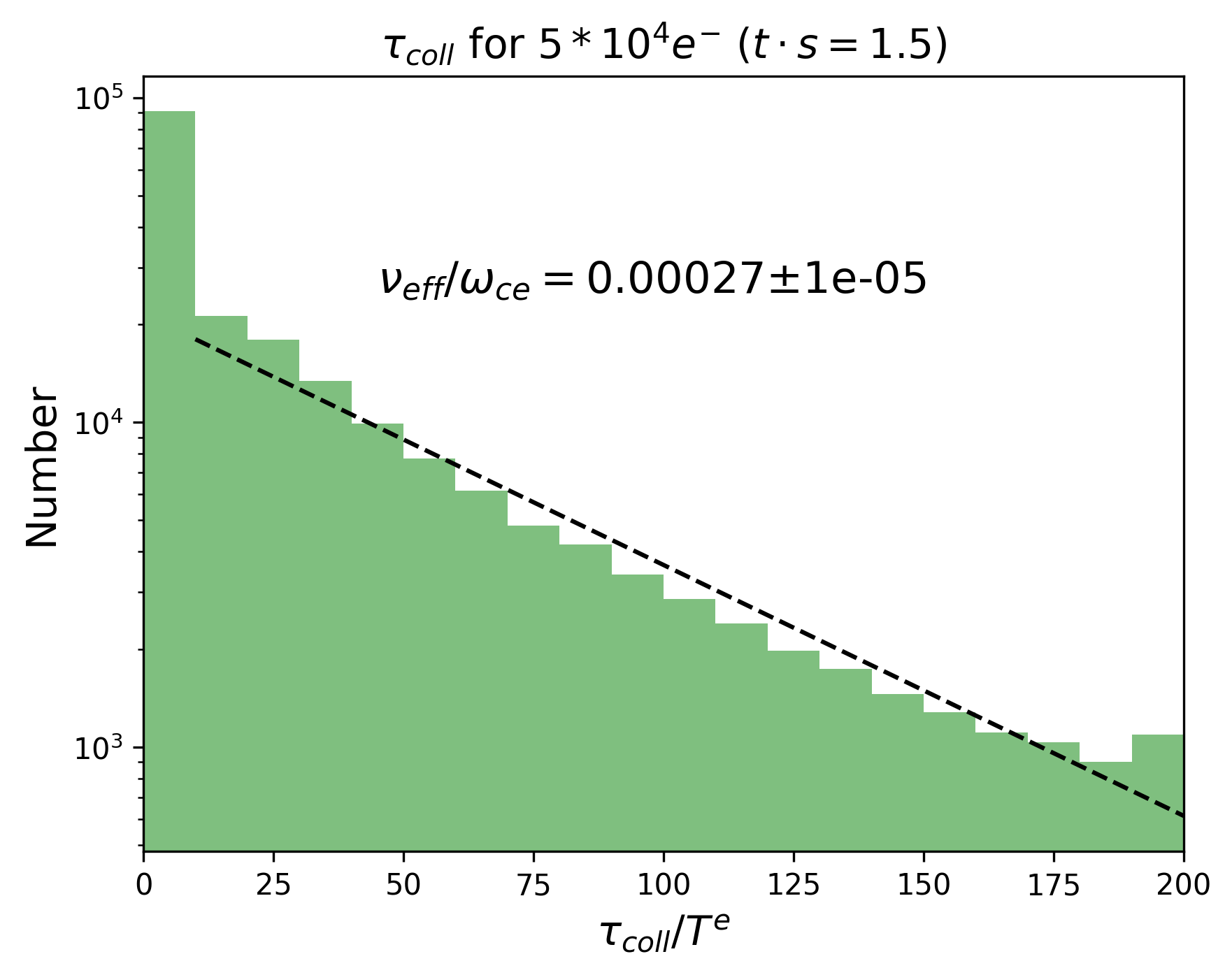}{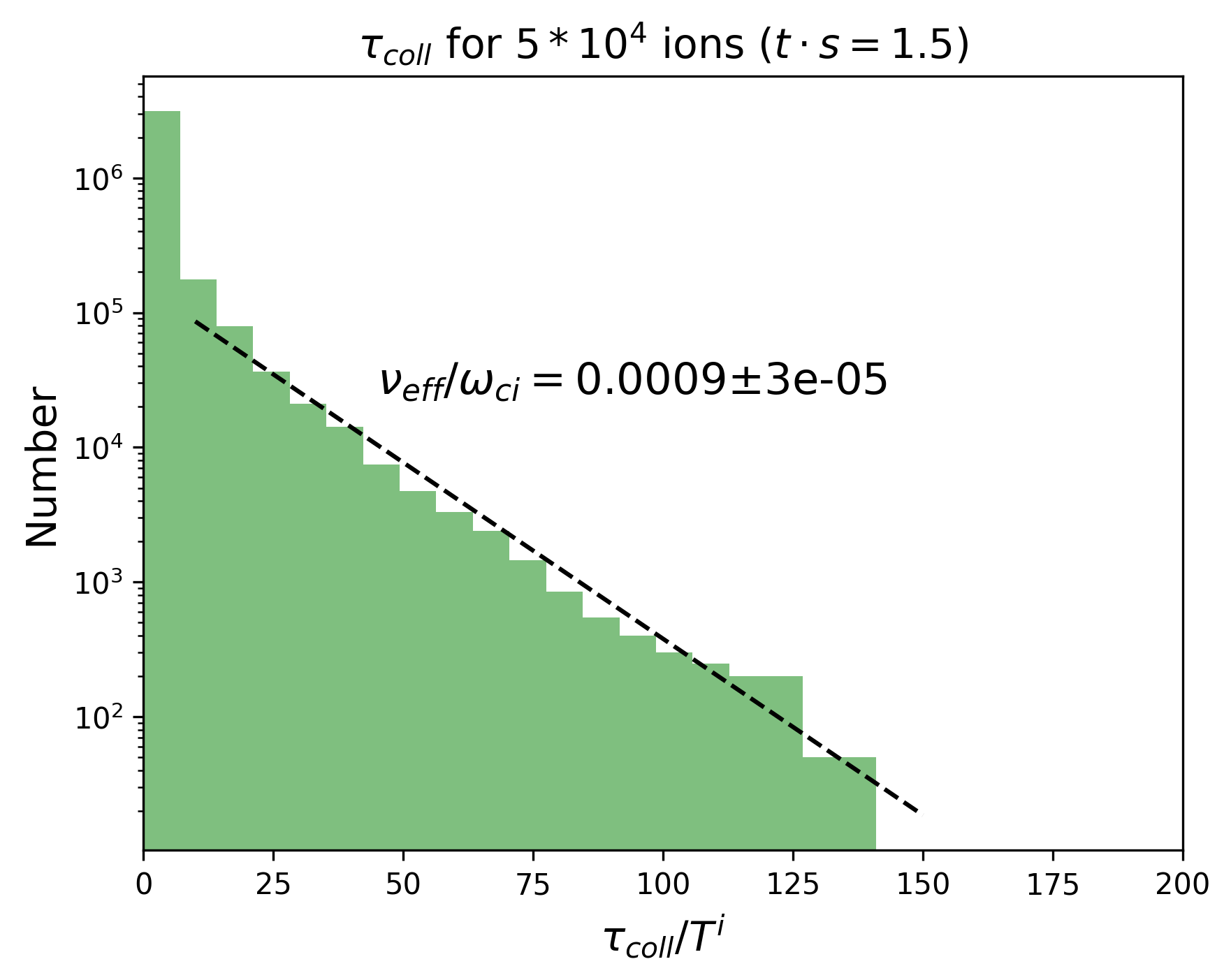}
\caption{Left: A histogram in log space showing the exponential fit of the bins of characteristic times of electrons to change their magnetic moment by a factor of $0.5e$. Right: A histogram in log space showing the exponential fit of the bins of characteristic times of ions to change their magnetic moment by a factor of $0.5e$.}
\label{fig:histograms}
\end{figure}

The slope of the fitted histogram provides an estimate of the scattering rate due to the whistlers and IC waves. An example of two fitted histograms of the characteristic times of both electrons and ions at $t \cdot s = 1.5$ can be seen in Figure~\ref{fig:histograms}.

\begin{figure}[h!]
\plottwo{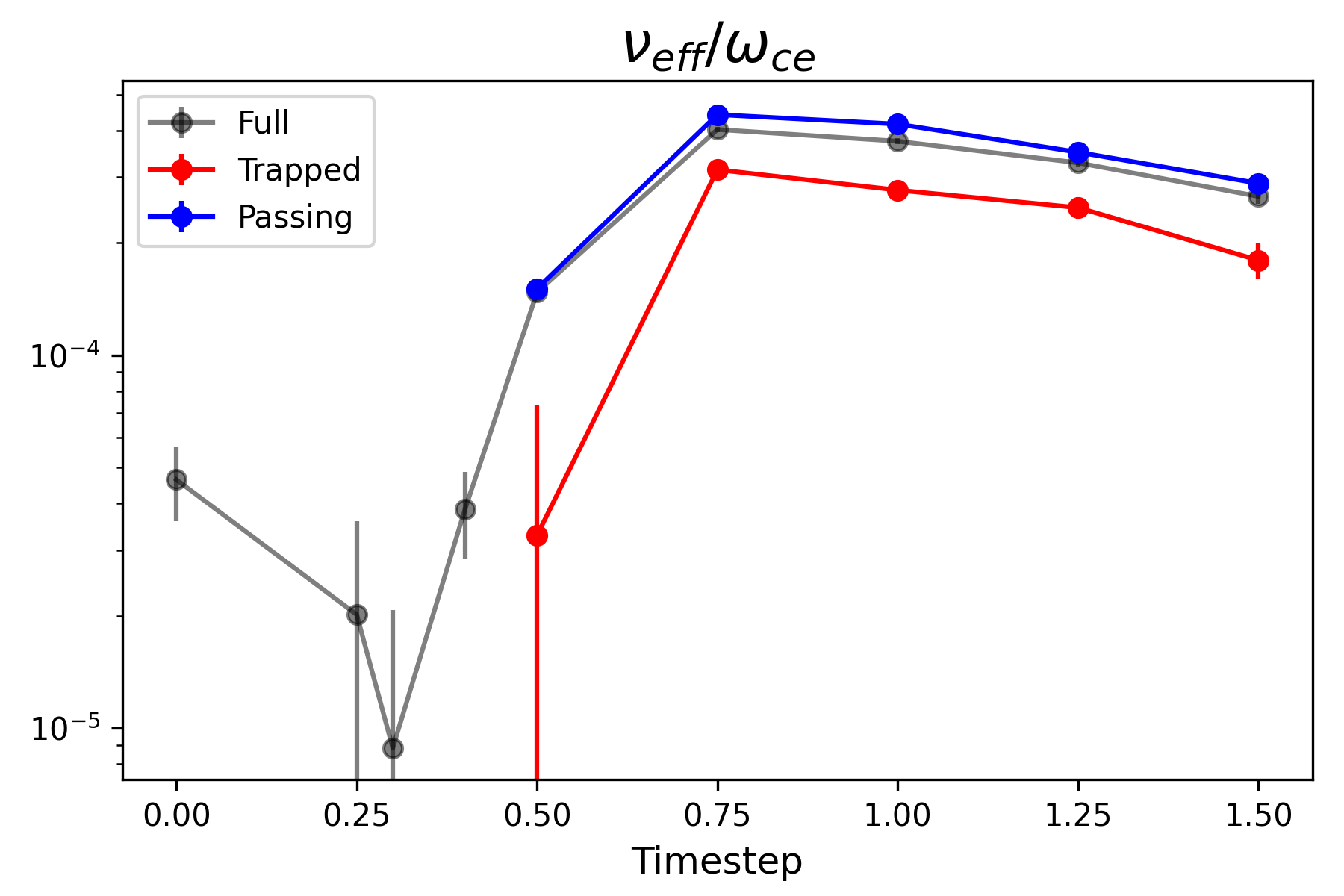}{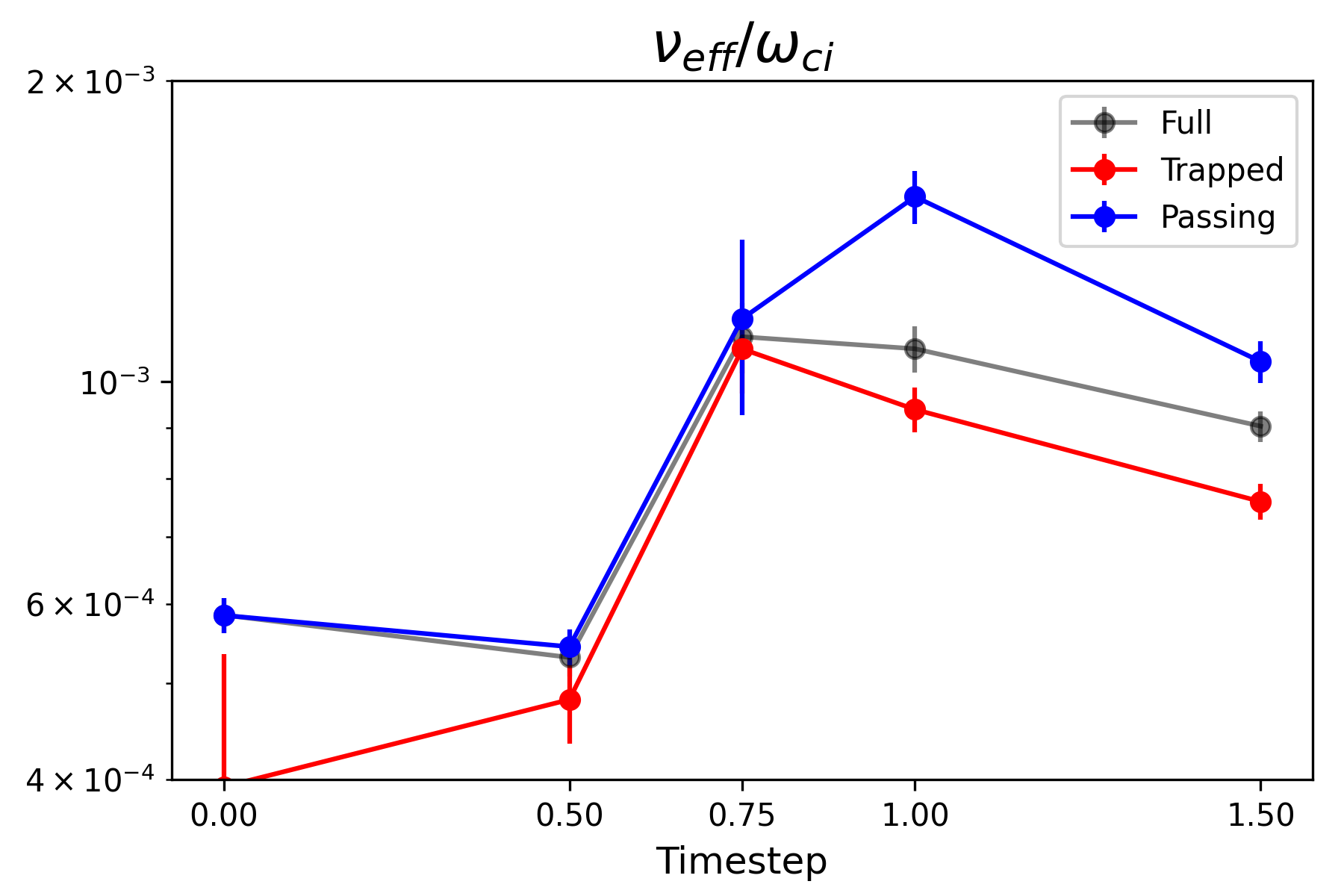}
\caption{These graphs in log space show the effective scattering rates of electrons/ions normalized by their respective gyrofrequencies over different timesteps separated into two populations of trapped and passing particles. }
\label{fig:scatteringRates}
\end{figure}

Now we can once again split the populations into trapped and passing particles based on the 50 sign changes criterion. We expect to see lower scattering rates for trapped particles, which, in order to be classified as ``trapped", have undergone less than 50 reversals of $v_{\parallel}$ in the chosen 200 gyroperiod interval. This is borne out by 
Figure~\ref{fig:scatteringRates}. Indeed, trapped particles have a clearly lower scattering rate than passing particles. 

For both electrons and ions, there is a considerable baseline collisionality level of $\nu_{\text{eff}}/\omega \sim 10^{-5}$  which is consistent between different mass ratios (Section~\ref{subsec:massRatio}), and an order of magnitude higher for ions. This can only be attributed to the PIC noise from the field initialization in TRISTAN-MP. Since this PIC noise should remain the same across snapshots, we can write $\nu = \nu_{\text{E}} + \nu_{\text{PIC}}$, where $\nu_{\text{E}}$ corresponds to the scattering rate only due to the scattering produced by growth in secondary waves (whistlers and IC waves) (\cite{Yerger2025}).

Importantly, there is a clear increase in the scattering rate that coincides with the growth of secondary instabilities. At $t\cdot s \approx 0.6$, we can see that the scattering rate increases for both ions and electron cases. This is exactly the time at which IC and whistler waves begin to grow (see $\delta B_z$,$\delta B_{\perp,xy}$ in Figure \ref{fig:fluctuations}), and also coincides with the end of the mirrors' secular growth (see $\delta B_{\parallel}$ in Figure \ref{fig:fluctuations}). This concurrence of events can be explained naturally from an energetics point of view. As secondary whistler and IC waves grow, they extract the free energy available in the pressure anisotropy built up by the trapped particles in regions of low magnetic field strength within mirror modes. In turn, this energy must be provided by the wave-particle interaction between secularly growing mirror modes and the particles which become trapped in the process, in such a way that there is a net energy transfer from mirror modes to trapped particles (\cite{SouthwoodKivelson1993}). As a consequence of this energy channeling, mirror modes can no longer increase their amplitude at the same rate as when trapping started, and mirror mode growth must slow down. This naturally explains the change in slope in the evolution of $\delta B_{\parallel}^2$ at $t\cdot s \approx 0.6$ in Figure \ref{fig:fluctuations}, at the exact time when the secondary waves start to grow (see rise in $\delta B_{xy,\perp}^2$ and $\delta B_z^2$ at $t\cdot s \approx 0.6$ in Figure \ref{fig:fluctuations}), marking the end of the mirror instability's secular stage and the beginning of its saturated stage, where mirror modes continue growing, but at a slower rate. This picture provides an interesting physical mechanism by which the mirror instability reaches saturation\footnote{This becomes most relevant in astrophysical applications where it is expected that kinetic instabilities like mirror reach their saturated state much faster than any relevant astrophysical timescales.}.

This effective collisionality enhancement indicates that the development of whistlers and ion-cyclotron waves plays an important role in pitch-angle scattering of the particles in the mirror saturated stage.

\subsection{Quasilinear Scattering}

In our simulations, the scattering of electrons by whistler waves is diffusive and follows the quasi-linear prediction \cite{KennelEngelmann1966}. In this regime, the scattering rate satisfies

\begin{align}
    \frac{\nu_{e, \text{eff}}}{\omega_{\text{ce}}} \sim \frac{\delta B_j^2}{B^2}.
\end{align}

Similarly, for ions scattering by ion cyclotron waves:

\begin{align}
    \frac{\nu_{i,\text{eff}}}{\omega_{\text{ci}}} \sim \frac{\delta B_j^2}{B^2}.
\end{align}

\begin{figure}[h]
    \centering
    \includegraphics[width=0.95\linewidth]{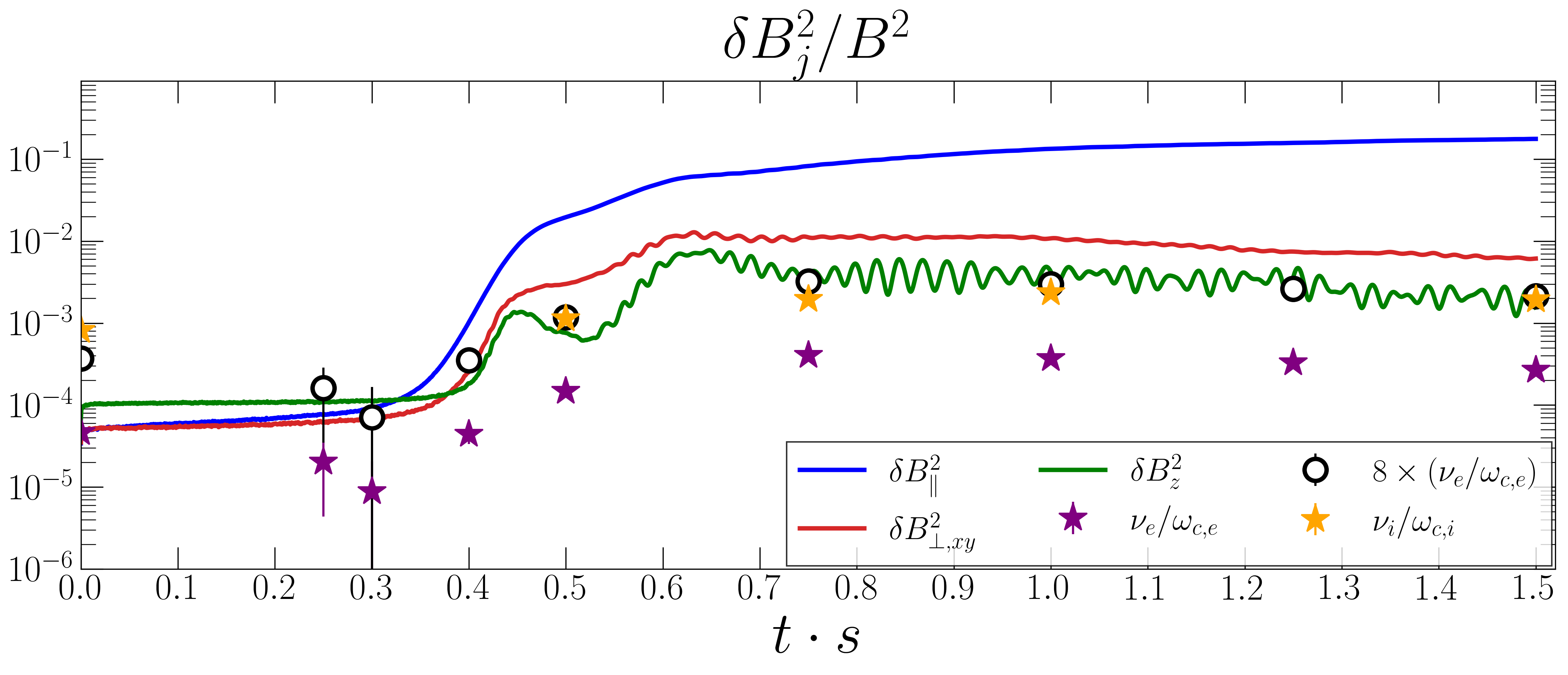}
    \caption{The evolution of the energy in the three component of the magnetic field fluctuations $\delta \textbf{B}$ normalized to $B(t)^2$ is shown: $\delta B_{\parallel}^2$ (blue line), $\delta B_{\perp,xy}^2$ (red line) and $\delta B_z^2$ (green line). The purple stars show the electron scattering rate normalized to the electron cyclotron frequency $\nu_{e,\text{eff}}/\omega_{\text{ce}}$. The open black circles show $\nu_{e, \text{eff}}/\omega_{\text{ce}}$ but amplified by an arbitrary factor of $8$. The orange stars show the ion scattering rate normalized to the ion cyclotron frequency $\nu_{i,\text{eff}}/\omega_{\text{ci}}$.}
    \label{fig:BFluctuations_Scattering}
\end{figure}

In our case, $j$ denotes the components of the magnetic field fluctuations perpendicular to \textbf{B}, i.e., $j=z$ or $j=\perp,xy$ (green and red line in fig. \ref{fig:fluctuations}$g$). We compare the scattering rate $\nu_{\text{eff}}$ obtained in our simulations with the evolution of the magnetic field fluctuations $\delta B_j/B$ in the TRISTAN simulations in figure \ref{fig:BFluctuations_Scattering}. We can see that the electron scattering rate $\nu_{e,\text{ eff}}/\omega_{\text{ce}}$ we obtained has a lower amplitude than $\delta B_z^2$ or $\delta B_{\perp,xy}^2$ (solid purple stars in fig. \ref{fig:BFluctuations_Scattering}), but it follows the same evolution. Indeed, if we multiply the electron scattering rate by a constant, arbitrary factor (in our case, a factor of $8$), we can easily see that it qualitatively follows the evolution of $\delta B_z^2$ (open black circles in fig. \ref{fig:BFluctuations_Scattering}). The ion scattering rate $\nu_{i,\text{eff}}/\omega_{\text{ci}}$ (orange stars in fig. \ref{fig:BFluctuations_Scattering}) matches up well with the 8x scaled electron scattering rate, following the same evolution as the magnetic field fluctuations. Note that any order-unity factor can be chosen arbitrarily, since this is a scaling argument with respect to the magnetic field fluctuations. 

\subsection{The effect of mass ratio} \label{subsec:massRatio}

In this section, we consider whether the results could depend on the mass ratio. The current state of computing resources makes achieving a realistic mass ratio of $m_i/m_e = 1836$ in PIC simulations unfeasible. Thus, most simulations run at much smaller mass ratios, from which it is important to consider the resulting non-physical effects. For example, it weakens the separation between ions and electrons, as ions could be more affected by electron-level effects.

\begin{figure}[h!]
\plotone{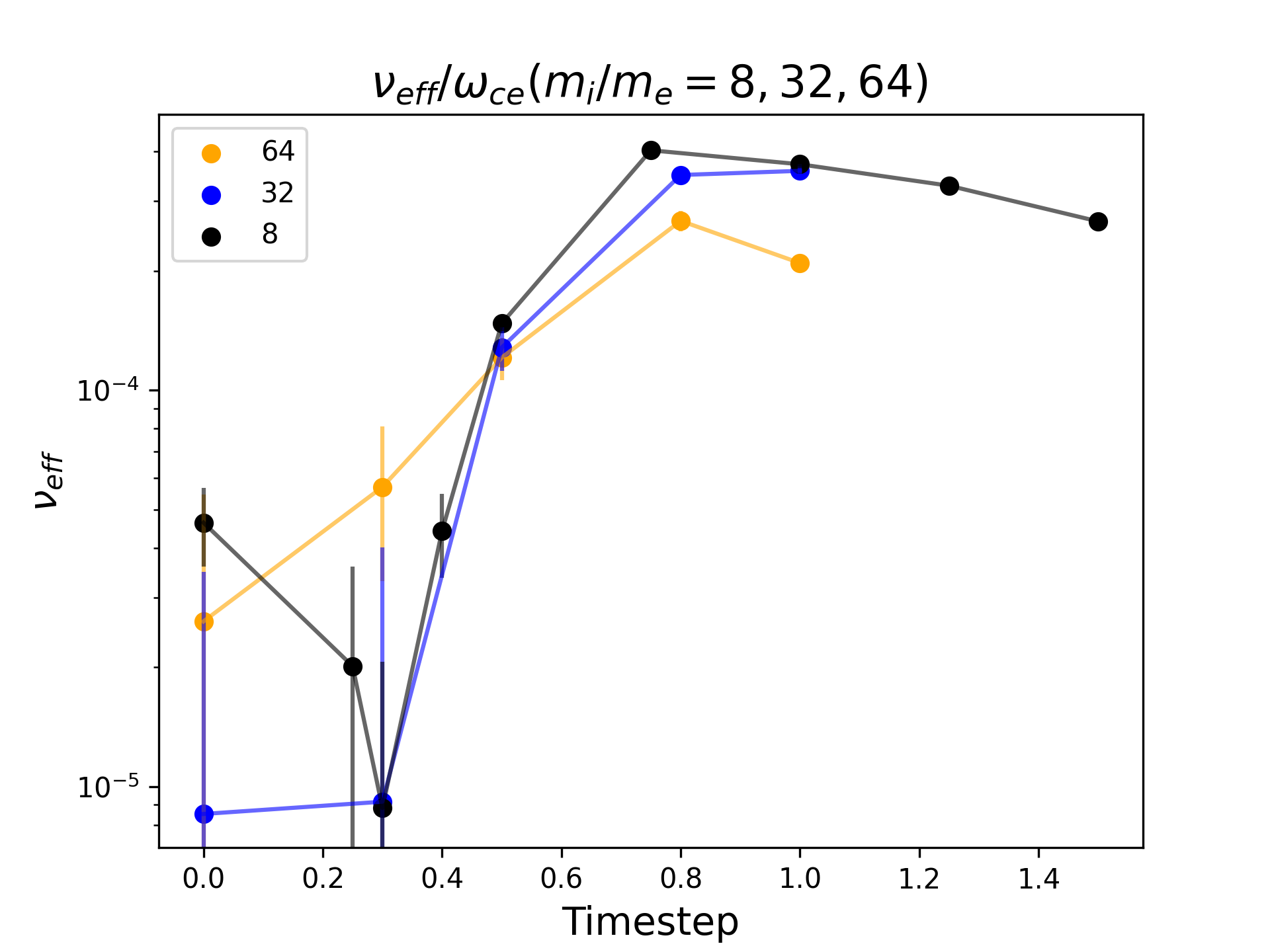}
\caption{the effective scattering rates of electrons normalized by their respective gyrofrequencies over different time steps, and separated with different mass ratios $m_i/m_e = 8$ (black line), $m_i/m_e = 32$ (blue line), and $m_i/m_e = 64$ (orange line).
\label{fig:massRatio}}
\end{figure}

However, this increase in scattering rates is consistent across higher mass ratios for electrons $(m_i/m_e = 8, 32, 64)$. This can be seen in Figure~\ref{fig:massRatio}. They all show a consistent rise in collisionality which levels out at around $t \cdot s = 1$. The scattering rates in all cases level off at similar 
collisionality levels of $\nu_{\text{eff}}/\omega_{c,e} \approx 0.0003$, with the lower rate at a mass ratio of 64 perhaps being attributable to the larger difference in whistler and ion cyclotron wave properties, which reduces the strength of electron interactions with ion cyclotron waves. This indicates that a lower mass ratio of 8 does not significantly impact the results of this work.

\section{Summary and Conclusions} \label{sec:conclusion}

In this work, we have implemented a particle propagation simulation to study the interaction of particles with mirror modes and the secondary waves generated by the mirror instability in the particle-in-cell simulations from \cite{Ley2023Secondary} (the TRISTAN simulations in this work). Taking snapshots of the TRISTAN simulation, the external fields are assumed to be static which is a good approximation given the interval of integration. After spawning in particles with random positions and velocities drawn from a Maxwell-Jüttner distribution with the same initial temperature as the corresponding TRISTAN simulations, we linearly interpolate the electromagnetic fields to the particle's position, and numerically integrate the orbits using the Boris method. Using a proxy variable of the number of times a particle's parallel velocity $v_\parallel$ changes sign to identify the trapping that mirror modes produce in ions and electrons, we split particles into populations of trapped and passing with trapped particles having more than 50 sign changes and passing particles having less than 50 sign changes (see Section~\ref{subsec: trapping}). Using our new classification of trapped and passing particles, we calculated the scattering rates of both particle populations and both species.

We found that the electric field fluctuations associated with all three instabilities have a significant effect on the scattering rates and on how the particle distribution functions evolve (see Figure \ref{fig:distfunctions}). Whereas dropping $\delta E$ leads to particle distributions with a clear loss cone, i.e. a deficit of trapped particles with predominantly parallel velocities (small magnetic moments), this is not seen in simulations which retain $\delta E$. We attribute this to energization of the particles due to the wave electric fields, an effect that is missing in the magnetic fluctuations only case. However, due to the relatively large wave phase velocities in our simulations dictated by computational resource considerations, the size of this effect may be overestimated (\S\ref{sec:simsetup}; penultimate paragraph), such that the simulations with and without electric fields bracket the effects wave-particle interactions under representative conditions.

We also  
found that there is a significant growth in the effective scattering rate of both ions and electrons, which is associated with the emergence of the secondary whistler and IC instabilities, and they provide the necessary scattering level to enhance the escape of trapped particle within mirror modes. The growth of the scattering rate coincides with the time at which IC and whistler waves are excited in the TRISTAN simulations ($t\cdot s \approx 0.6$), which also marks the end of the secular phase of the mirror instability. We argue that this is not accidental. From an energetics point of view, the secondary whistler and IC waves grow by extracting energy from the pressure anisotropy built up by the particles which undergo trapping within mirror modes. Therefore, mirror modes must provide the energy via wave-particle interaction with the trapped population. As this energy channels into whistler and IC wave growth, it can no longer be used for increasing mirror modes amplitude, and therefore the rate of mirror mode growth slows down. This naturally explains the change in slope in $\delta B_{\parallel}^2 (t)$, which marks the beginning of the saturated stage of the mirror instability (see fig. \ref{fig:fluctuations}). The excitation of these secondary waves is relevant for regulating the global anisotropy of ions and electrons in the plasma driving it towards marginal stability of the ion-cyclotron instability for ions, and whistler instability for electrons\footnote{In the case of electrons, it is less clear that their pressure anisotropy gets pinned to the whistler threshold (see \cite{Ley2023Secondary}). This could be a hint for a much richer dynamics at very late stages of the mirror instability (see e.g. \cite{Jiang2022})}
(\cite{Ley2023Secondary}). Importantly, we found that the electron scattering rates closely track quasilinear scaling (Figure \ref{fig:BFluctuations_Scattering}).

The viscous heating via anisotropic viscosity produced by turbulent motions in the ICM is ultimately controlled by the level of pressure anisotropy that the plasma reaches when kinetic microinstabilities have saturated (\cite{Lyutikov2007, Kunz2011, Ley2023}). As we saw, the interaction of the secondary IC and whistler waves with trapped ions and electrons, respectively, enhances particle escape and becomes essential for the saturation of mirror modes. During this process, it is observed that the ion pressure anisotropy is pinned to the ion-cyclotron threshold instead of the mirror threshold (\cite{Ley2023Secondary}), the IC threshold being higher by a factor of ${\beta_i^{1/2}}$. This results in an enhancement of the heating efficiency by the same amount. The consequences this enhancement can have on the thermal stability of the heating mechanism might be worth exploring. The presence of secondary whistler waves can also affect the onset and evolution of other types of instabilities, namely, the heat flux driven whistler instability, which could influence the heat conduction in galaxy clusters (\cite{Yerger2025}). 

A limitation of this work has been the collisional baseline in the scattering rate measurements which seems intrinsic to the underlying PIC simulation noise. Future work would require new simulations with a higher number of macroparticles per cell, $N_{\text{ppc}}$, and potentially even a higher mass ratio to advance understanding of the nonlinear stages of the mirror instability and the interaction of whistler/IC waves with the magnetic trapping process.

\section*{Acknowledgments}

We thank Karol Fulat for helpful advice on filtering the electric and magnetic fields to reduce PIC noise. We thank Aaron Tran and Daniel Humphrey for useful discussions.
Francisco Ley acknowledges support from NSF grant No. PHY-2010189. 
Petr Ugarov acknowledges support from NSF grant No. PHY-2409316. Ellen Zweibel acknowledges the support of NSF grant PHY-2409224 and from the Simons Foundation through the Simons Collaboration on Extreme Electrodynamics in Compact Sources, as well as the hospitality of the Leonard Parker Center for Cosmology and Gravitational Astrophysics at U. Wisconsin-Milwaukee, where some of this work was completed. This research was performed
using the computing resources and assistance of the UW-
Madison Center For High Throughput Computing (CHTC) in
the Department of Computer Sciences. 

\appendix

\section{Energy Conservation and Heating} \label{sec:appendix}

\begin{figure}[h!]

\plottwo{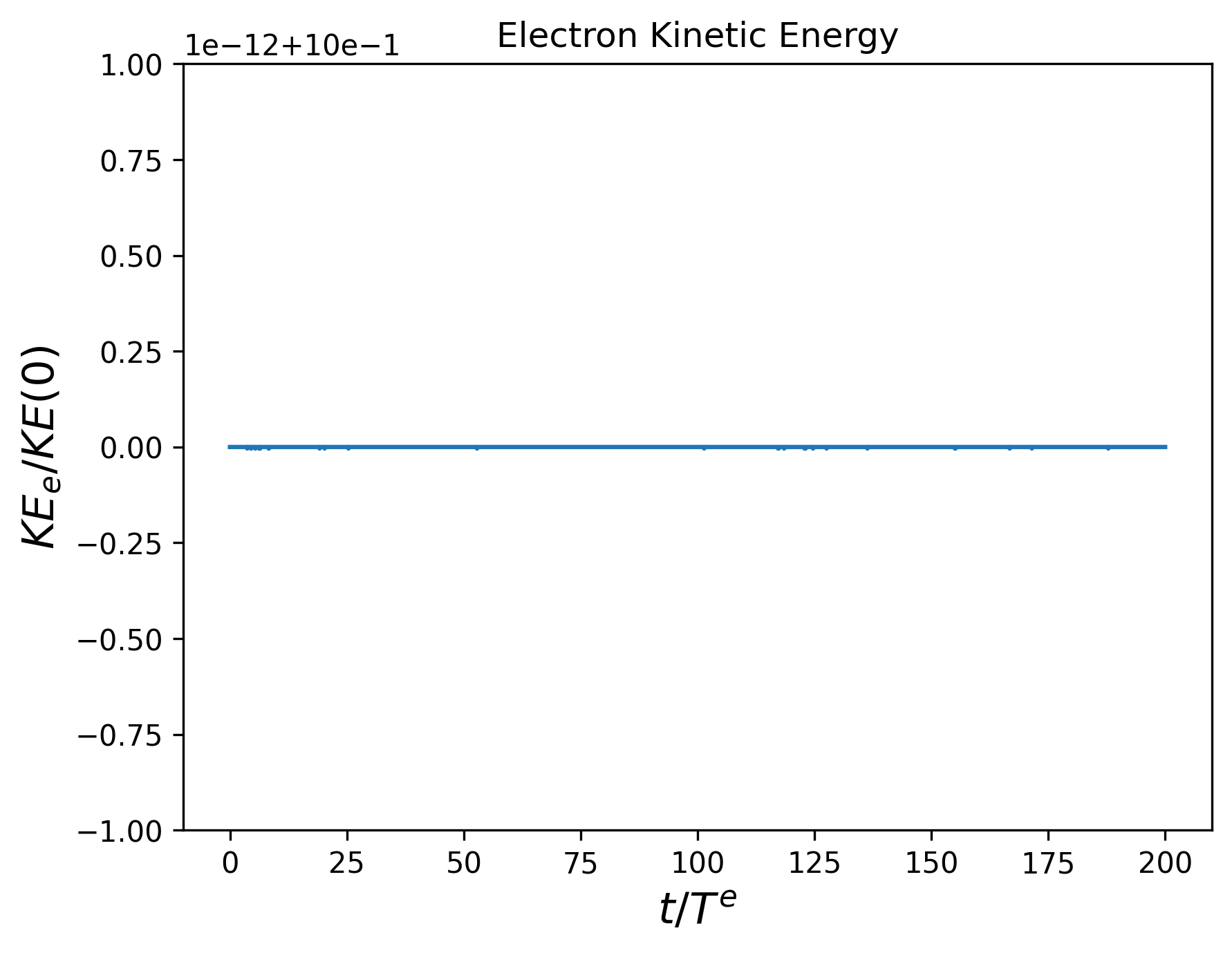}{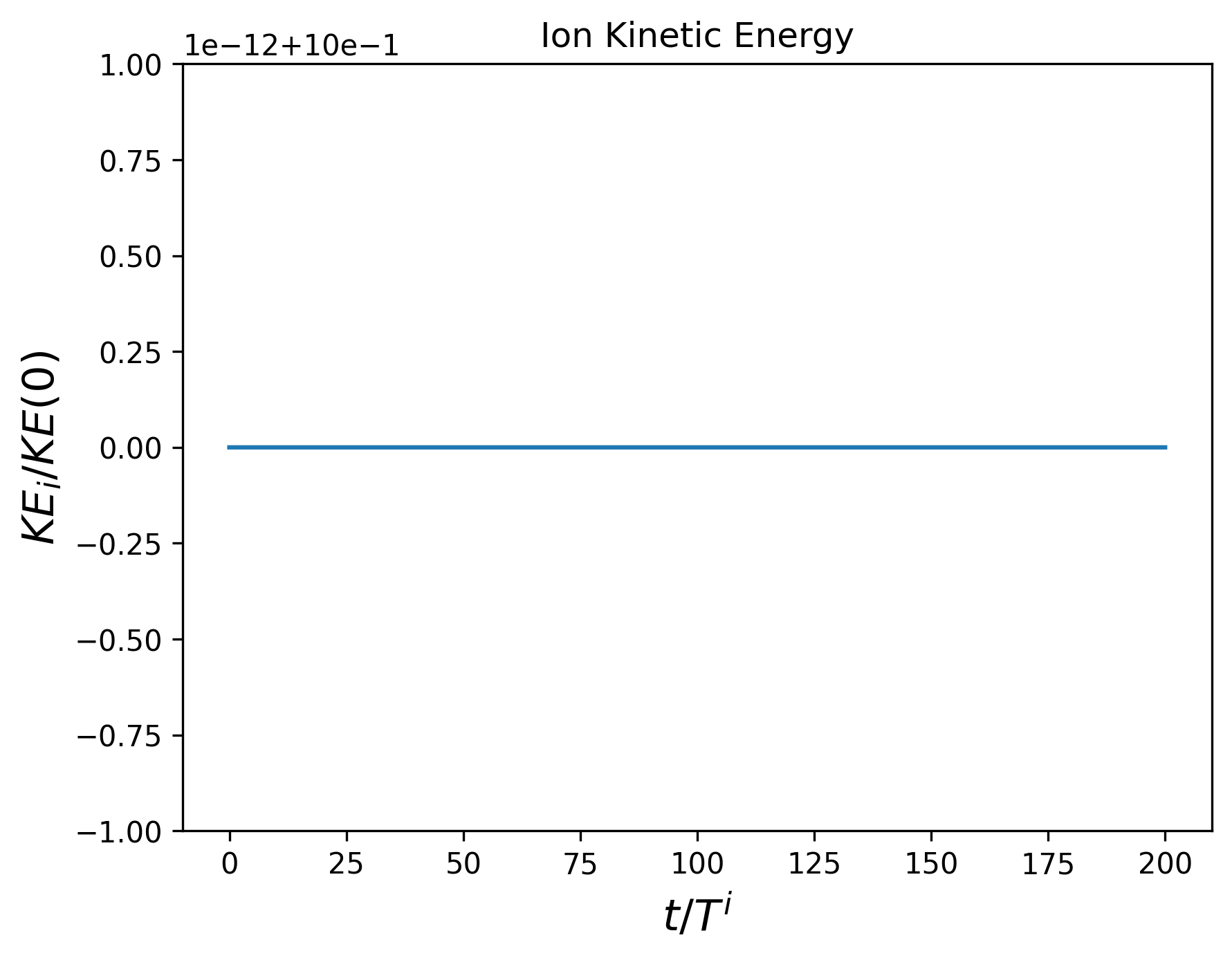}
\caption{These plots show the kinetic energy stability over 200 ion/electron gyroperiods of an ion or electron without electric fields at $t \cdot s = 1.5$. Note that the y-axis shows variations from 1 as the ratio of kinetic energy to the initial kinetic energy of the particle. 
\label{fig:KEComparison}} 

\end{figure}

We find that the simulation at these numerical parameters has good numerical convergence. As can be seen in Figure \ref{fig:KEComparison}, the simulation conserves energy up to less than $10^{-12}$ without electric fields, ensuring that both the numerical integration and field interpolation have low error, as expected for the Boris method. 

\begin{figure}[h!]

\plotone{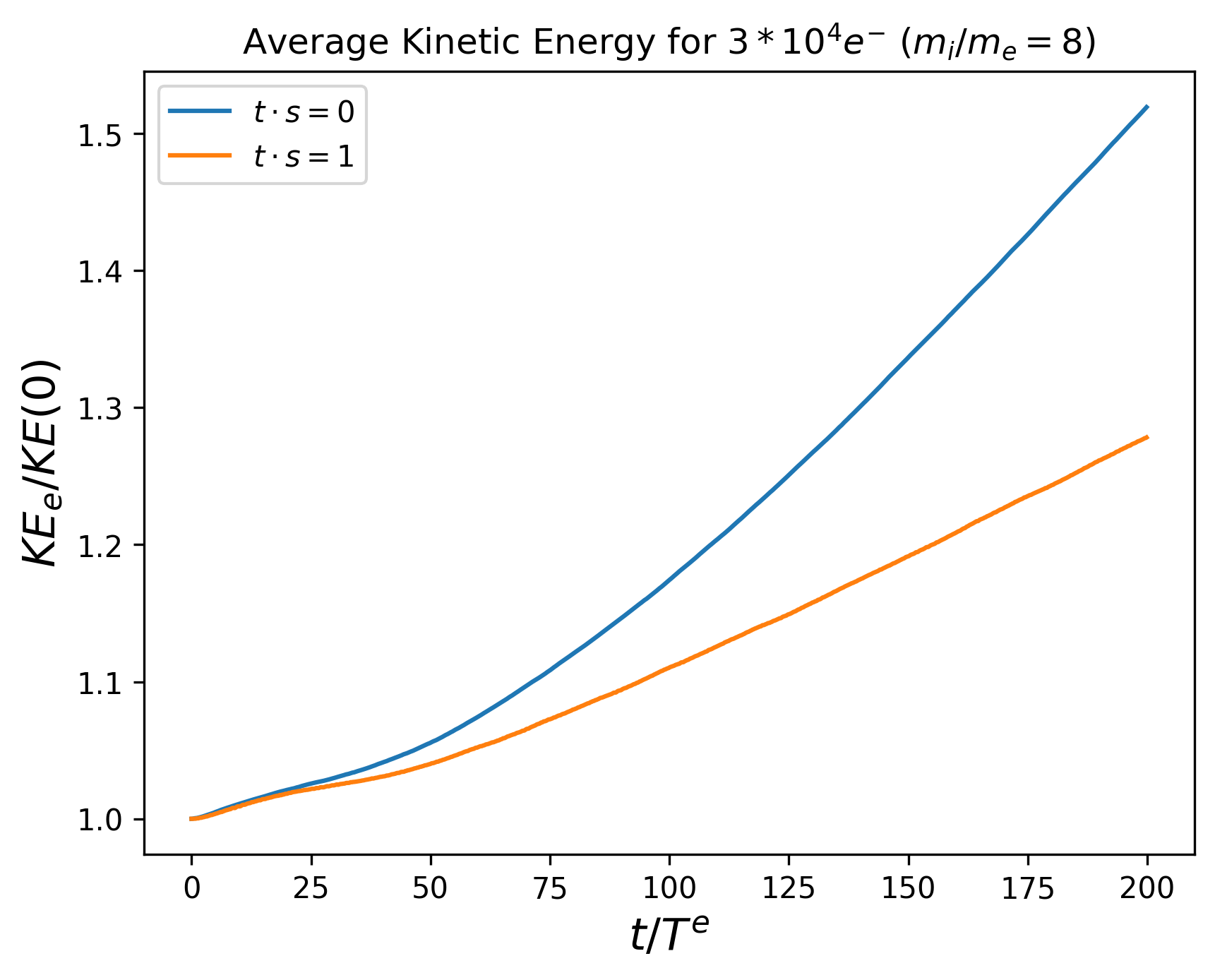}
\caption{Average kinetic energy normalized by initial kinetic energy for $3\times 10^4$ electrons integrated over 200 electron gyroperiods for different timesteps $(t \cdot s = 0, 1).$
\label{fig:heating}}

\end{figure}

However, when electric fields are added, the particles can however exhibit noticeable heating and energy gain as can be seen in Figure~\ref{fig:heating}. The heating was significantly reduced by filtering the magnetic and electric fields using a Gaussian filter at the cutoff of one electron Larmor radius to keep physically relevant spectral power in ion cyclotron and whistler modes and reduce PIC noise which is primarily concentrated in higher wavenumbers $kR_{L,e} \gtrsim 1$. Figure~\ref{fig:heating} shows the kinetic energy gain of electrons at the different timesteps $t \cdot s = 0, 1$. We can see that the energy gain is consistent across different snapshots with the static snapshot of $t \cdot s = 0$ having the highest energy gain even in the absence of any instabilities. This indicates that the heating can be largely attributed to electric field fluctuations from PIC noise. Even so, there is some minor difference between the heating rates at different snapshots. We hypothesize that this is due to the interplay of mirror modes and secondary instabilities. In the $t \cdot s = 1.0$ snapshot the mirror instability is saturated, which traps particles. Consequently this makes them interact less with electric field fluctuations from PIC noise throughout the simulation box, and so they experience less heating. We expect that heating rates for ions would be similar.

\bibliography{main}{}
\bibliographystyle{aasjournal}

\end{document}